\documentclass[letterpaper,11pt]{article}
\usepackage{jheppub2} 
\usepackage[T1]{fontenc}

\usepackage{natbib}
\usepackage{graphicx}
\usepackage{hyperref}
\usepackage{amsmath}
\usepackage{amsfonts}
\usepackage{amssymb}
\usepackage{enumitem}
\usepackage{tocvsec2}
\usepackage{subfigure}

\newcommand{\ZZ}{{\mathbb Z}}

\def\tr{\text{tr}\,}

\hypersetup{
    colorlinks=true,   
    linkcolor=red,     
    citecolor=blue,    
    filecolor=magenta, 
    urlcolor=blue      
}

\makeatletter
\def\l@subsubsection#1#2{}
\makeatother    

\advance\footnotesep 1.5pt

\begin{document}

\title{Lifetimes of (near) eternal false vacua}

\author[]{Aleksey Cherman}
\emailAdd{acherman@umn.edu}

\author[1]{and Theodore Jacobson}
\emailAdd{jaco2585@umn.edu}
\affiliation[1]{School of Physics and Astronomy, University of Minnesota, Minneapolis MN 55455, USA}

\abstract{We consider examples of long-lived false vacua in quantum
field theory that arise from so-called `universes'. These false vacua are
protected by a $(d-1)$-form global symmetry, where $d$ is the
dimension of spacetime. The lifetimes of the false vacua are set by UV data: the
tension of $(d-2)$-branes charged under a $(d-2)$-form gauge symmetry. The
lifetimes can be made parametrically long even when the difference in energy
density between the false and true vacua is large compared to the natural scales
of the field theory. We study examples of near-eternal false vacua in abelian gauge
theories in two dimensions and in four-dimensional QCD.  In both cases, it is
possible to view the $(d-1)$-form symmetries as arising from a modification of
the sum over instantons.  We find that the modification of the instanton sum in
4d QCD leads to a higher-group symmetry structure involving the 3-form and
conventional 0-form global symmetries.} 

\maketitle

\maxtocdepth{subsection} 
 
\section{Introduction}
\label{sec:intro} 

False vacua are commonplace in QFT.  But we are used
to the idea that if we wait long enough, the system always ends up in the true
vacuum~\cite{Kobzarev:1974cp,PhysRevD.9.2291,PhysRevD.14.3568,STONE1977186,Coleman:1977py,Callan:1977pt}.
Indeed, suppose we prepare a system in a false vacuum state.  In familiar
situations, false vacuum decay proceeds by bubble nucleation:  there is a
non-zero probability for a bubble of the true vacuum to appear thanks to quantum
or thermal fluctuations.  Heuristically, small bubbles collapse to zero size due
to the dominance of the bubble wall tension in their energy, but if a large enough bubble of the
true vacuum appears, then the system can lower its energy by making the bubble
bigger and bigger due to the difference in energy densities between the true and
false vacua.  As a result, sufficiently large bubbles expand and convert the
entire system to the true vacuum. The lifetime of the false vacuum decreases as
the difference in energy densities between the false and true vacuum is
increased.

Taken at face value, the discussion above might suggest that the notion of
`eternal false vacua' in the title of this paper is an oxymoron: a false
vacuum should never be eternal in QFT. However, some recent theoretical
developments imply that in fact eternal false vacua do exist in some QFTs. The
examples we will study the following three properties:\footnote{These properties
are sufficient but not necessary for the existence of universes. For instance,
in 2d Maxwell theory the existence of a $U(1)$ electric $1$-form global symmetry
alone gives rise to eternal false vacua characterized by expectation values of
the local operator $\star F = \frac{1}{2}\epsilon^{\mu\nu}F_{\mu\nu}$, which is
nothing but a constant electric field. }
\begin{enumerate}
    \item a $\ZZ_{p}$ $(d-1)$-form global symmetry, denoted by $\ZZ^{(d-1)}_{p}$.
    \item some point in its parameter space where a global 0-form symmetry $G^{(0)}$ appears.
    \item a mixed 't Hooft anomaly involving $G^{(0)}$ and $\ZZ^{(d-1)}_{p}$.
\end{enumerate}
The first (and consequently third)  properties above may seem rather exotic, but
we will see that there are some very simple examples of theories with these properties.  

In our examples $G^{(0)}$ will be an Abelian group.  At the point in parameter
space where $G^{(0)}$ is a symmetry, the long-distance effective field theory
cannot be trivial due to the 't Hooft anomaly, and in the examples we will
discuss $G^{(0)}$ is spontaneously broken, so that there are $|G^{(0)}|$
degenerate vacua. However, another consequence of the `t Hooft anomaly is that
the would-be domain walls connecting these vacua have \emph{infinite} tension.
So even when the spatial volume is finite, the degenerate vacua cannot mix, in
sharp contrast to more familiar examples of QFTs with spontaneously broken
symmetries.  We will follow 
\cite{ZoharPrivateCommunications,Tanizaki:2019rbk,Komargodski:2020mxz}
 and call distinct vacua with the peculiar property above
 `universes.'  Indeed, so long as $\ZZ^{(d-1)}_{p}$ is not explicitly broken,
 no local measurement within a given universe can tell us anything about the
 other universes - including whether they even exist!  Only the
 behavior of $(d-1)$-dimensional extended operators can probe
 the differences between the universes.  More physically, if the
 $\ZZ^{(d-1)}_{p}$ global symmetry is slightly broken, then the 
 `disconnected' universes all get connected into one universe.  The fact that a
 given QFT gives rise to distinct universes in the limit that $\ZZ^{(d-1)}_{p}$
 becomes exact has important implications even when $\ZZ^{(d-1)}_{p}$ is slightly
 broken so that there is only one universe, which we will explore here.
 
Let us first suppose $\ZZ^{(d-1)}_{p}$ is not broken explicitly.  Then we can
ask what happens if we explicitly break the global symmetry $G^{(0)}$ by dialing
parameters in the QFT.  Clearly the  Poincar\'e-invariant local minima of the
quantum effective potential become non-degenerate, leading to the appearance of
a unique minimum-energy vacuum. Normally, this would imply that a locally-stable
vacuum with a non-minimal value of the energy density --- a false vacuum ---
would decay to the true vacuum in the thermodynamic equilibrium limit. But here
this decay is impossible, and the lifetime of the false vacuum is infinite. This
should be viewed as a remnant consequence of the 't Hooft anomaly between
$G^{(0)}$ and $\ZZ^{(d-1)}_{p}$. Bubbles of any finite size simply cannot
appear, because the surface tension of such a bubble is infinite.  To get a
finite false-vacuum lifetime, one must also explicitly break the
$\ZZ^{(d-1)}_{p}$ symmetry.  Breaking the $\ZZ^{(d-1)}_{p}$ symmetry entails
coupling the system to dynamical $(d-2)$-branes: objects with a
$(d-1)$-dimensional worldvolume with a tension $T_{d-2}$.  The wall tension of a
bubble of true vacuum is  $\sim T_{d-2}$.  As a result, the false vacuum
lifetime is determined by the exponential of the ratio of the tension $T_{d-2}$
of these branes and the dimensionful scales of the original QFT.

QFTs with unbroken $n$-form global symmetries are believed to be inconsistent
with quantum gravity constraints (such QFTs are termed to be `in the
Swampland'), see e.g. Ref.~\cite{Palti:2019pca}.   So for phenomenological
applications of these ideas, one should assume the $(d-2)$-branes that are
sources for the $\ZZ^{(d-1)}_{p}$ symmetry have a finite tension.  This tension
can be extremely large ($T_{d-2} \sim M_{\rm Planck}^{d-1}$) compared to the
energy scales which are native to our $d$ dimensional QFT, leading to extremely
long-lived false vacua, without any need to tune the parameters controlling the
low-energy physics.

In what follows we will do a warm up in quantum mechanics, and then discuss some
simple examples of universes in QFTs in $d= 2$ and $d=4$.   In $d=2$ we will
focus on variants  of the venerable Schwinger model:  that is, quantum
electrodynamics in $d=2$. In this simple example, there are two ways to
interpret the $(d-1)$-form symmetry.  One way to obtain the $1$-form symmetry is
to assume the matter fields have charge $p$, and then the $\ZZ^{(d-1)}_{p}$
symmetry is just the $1$-form `center' symmetry.  The other way to obtain the
1-form symmetry is to have matter with charge $1$, but modify the instanton sum
to only include instantons with topological charge divisible by $p$.  In  either
case, $G^{(0)}$ is a discrete chiral symmetry.  The $1$-form symmetry can be
broken by adding massive matter fields with appropriate gauge charges.   The 2d
analysis can be done very explicitly using 2d bosonization techniques. 

After that we make the jump to $d=4$, where we study $SU(N)$ QCD with $N_f$
fundamental fermions and a modified instanton sum, generalizing the analysis
of~\cite{Tanizaki:2019rbk}.  In this variant of QCD the instanton sum is
modified such that only instantons with topological charge divisible by a
positive integer $p$ are allowed to contribute to the path integral.  Then
$G^{(0)} = \ZZ^{(0)}_{2N_fp}$ is a discrete chiral symmetry, while
$\ZZ^{(d-1)}_{p}$ is a $3$-form symmetry when $d=4$.   We show that the $3$-form
symmetry and the 0-form continuous symmetry groups of QCD do not form a direct
product. Instead, they mix non-trivially so that the symmetry group of QCD with
a modified instanton sum should be thought of as a 4-group.  Breaking the
higher-form symmetry requires coupling our variant of 4d QCD to dynamical
$2$-branes. The hierarchy between the scale set by the 2-brane tension $T_2$ and the QCD scale $\Lambda_{\text{QCD}}$ implies that false vacua in the modified QCD theory are very long lived.

\section{Universes in quantum mechanics}
\label{sec:qm} 

\subsection{`Eternal false vacua' are trivial in quantum mechanics}
\label{sec:trivial} 

In quantum mechanics, we are used to the idea that excited states cannot decay
provided they are exact eigenstates of the Hamiltonian. If the Hamiltonian
enjoys some global symmetry $G$, then there are no transitions between an
excited state carrying global charge to a lower energy zero charge state such as
the vacuum. In this sense, the notion of an `eternal false vacuum' is somewhat
trivial in quantum mechanics. 

For instance, consider a particle in a symmetric double-well potential. The
spectrum organizes itself into representations of parity, which acts as $x \to
-x$. The ground state is the symmetric combination of node-less wave functions
localized at the two wells, and is invariant under parity. The first excited
state is the anti-symmetric combination, which is odd under parity. On top of
each of these states is a tower of excitations, each being even or odd under
parity. We may view each tower as being distinct universes, in the sense
that they are protected by a selection rule: only parity odd operators can
connect states between the two sectors.  (Of course, the excited states within
each tower cannot decay either, since they are energy eigenstates.)

Going back to generalities, suppose the system has a second global symmetry $G'$
which has a mixed 't Hooft anomaly with $G$. As a consequence, the ground state
can not be unique. We can choose a basis such that the lowest energy states are
eigenstates of $G$, but not $G'$ (they form a $G'$ multiplet). In this way, the
symmetry $G'$, together with the 't Hooft anomaly, ensure that the universes
associated with $G$ are degenerate. If we break $G'$ explicitly this degeneracy
is lost, though the universes remain. 

An example of the above situation is furnished by a free particle on a circle
with a $\theta$-angle~\cite{Gaiotto:2017yup,Kikuchi:2017pcp,Aitken:2018kky}. At
$\theta =\pi$ there is a 't Hooft anomaly between the $U(1)$ shift symmetry and
charge conjugation, which together comprise an $O(2)$ symmetry. This symmetry is
projectively realized on the Hilbert space. The spectrum consists of states
$|n\rangle$ labeled by their $U(1)$ charge. According to the discussion above,
each state is its own universe. The two lowest-energy universes $|0\rangle$ and
$|1\rangle$ are degenerate and are exchanged by charge conjugation. Once we
break charge conjugation by moving away from $\theta =\pi$, one of these states
becomes the true vacuum. However, the $U(1)$ symmetry prohibits the decay of the
higher-energy states to this unique vacuum. 

The discussion so far serves to show that in quantum mechanics, universes and
eternal false vacua are overly-complicated ways of thinking about completely
standard concepts. However, it will be useful to consider an additional
quantum-mechanical example in more detail. The following model exhibits many of
the features of the quantum field theories discussed later in the paper. 

\subsection{A quantum mechanics warmup}
\label{sec:qm_example} 

Consider a quantum mechanical system consisting of three particles with
coordinates are $x,y,z$ respectively on a circle of radius $2\pi L$ with
Euclidean-time Lagrangian 
\begin{equation}
\mathcal L =  \frac{1}{2g}\left(\dot x^2 + \dot y^2\right) + \frac{i}{2\pi}x \, \dot y +\frac{i}{2\pi}\chi (\dot y-p\, \dot z) + V(x,y,z) \, , \label{eq:qmexample} 
\end{equation}
where we took units where $L=1$ and $p \in \ZZ_{\ge 0}$.\footnote{This model is
related to the small-circle limit of the charge-$p$ Schwinger model on
$\mathbb{R}\times S^1$ which appears in~\cite{Komargodski:2017dmc}. For the
precise connection between two-dimensional abelian theories with non-minimal
charge and modified instanton sums, see Sec.~\ref{sec:modified_sum}.} We assume
that the $x$ and $z$ periodicities in the potential are $2\pi/p$ and the $y$ periodicity in the potential is
$2\pi$.  Although the variables $\chi$ and $z$ do not have
kinetic terms, they are dynamical in the sense that we integrate over all
configurations of $\chi$ and $z$ in the path integral. The $2\pi$-periodic
variable $\chi$ plays the role of a Lagrange multiplier setting $\dot y = p \,
\dot z$. Consequently, winding numbers of $y$ are constrained to be multiples of
$p$, 
\begin{equation}
\int dy = p \int dz \in 2\pi p\, \mathbb{Z}. 
\end{equation}
To keep things simple, in the following we will set $V(x,y,z) = V(x)$ below
unless specified otherwise. This allows us to demonstrate
many features exactly (by integrating out $z$ explicitly, for instance) rather than making general statements and then checking
them in perturbation theory. 

The equation of motion for $z$ sets $\dot\chi=0$,
and summing over winding configurations with $\int dz \in 2\pi \ZZ$ sets $\chi =
2\pi m/p$ for $m\in\ZZ$. The variable $e^{i\chi}$ is therefore $\ZZ_p$-valued,
and summing over its discrete values enforces the constraint $\int dy \in 2\pi p
\,\ZZ$ as a delta function in the path integral.\footnote{This constraint on the
winding numbers of $y$ still holds when $\chi$ and $z$ have kinetic terms,
despite the fact that $\chi$ is no longer a discrete-valued field in this case.
For simplicity we set these kinetic terms to zero. } 

For our purposes the relevant symmetries of the model are the two discrete
$\ZZ_p$ shift symmetries
\begin{equation}
(\ZZ_p)_x: x \to x + \frac{2\pi}{p}, \, \chi\to\chi-\frac{2\pi}{p}, \quad (\ZZ_p)_z : z \to z + \frac{2\pi}{p} \, .
\end{equation}
The symmetry operator of $(\ZZ_p)_z$ is $e^{-i\chi}$, in the sense that 
\begin{align}
    \langle
e^{-i\chi(t_0)}\, e^{iz(t_1)}\rangle = e^{\frac{2\pi i}{p}\Theta(t_1-t_0)}\,
\langle e^{iz(t_1)}\rangle \, .
\label{eq:QM_z_sym_gen} 
\end{align} 
To verify this relation one can  write
\begin{align}
\langle e^{- i\chi(t_0)} e^{iz(t_1)} \rangle =  \frac{1}{Z} \int 
\mathcal{D} x\, \mathcal{D} y\, \mathcal{D} z\, \mathcal{D}\chi\, e^{-(S +\delta S)} 
e^{iz(t_1)}
\end{align}
where $\delta S = i \int dt\, \chi(t) \,\delta(t_0)$.  We can cancel this shift in the
action by a shift of $z(t) \to z(t) + \frac{2\pi}{p} \Theta(t-t_0)$, which leads to
\begin{align}
    \langle e^{- i\chi(t_0)} e^{iz(t_1)} \rangle = \frac{e^{i \frac{2\pi}{p}\Theta(t_1-t_0)}}{Z}
    \int 
    \mathcal{D} x\, \mathcal{D} y\, \mathcal{D} z\, \mathcal{D}\chi\, e^{-S} 
    e^{iz(t_1)}\,,
\end{align}    
reproducing  \eqref{eq:QM_z_sym_gen}.  Since $e^{-i\chi}$ itself is charged
under $(\ZZ_p)_x$, these two shift symmetries have a mixed `t Hooft
anomaly~\cite{Gaiotto:2014kfa}. The arguments summarized in the introduction
suggest that this ought to give rise to $p$ universes. 

To see these distinct universes explicitly, it is convenient to integrate out
$y$ and $z$ to isolate the dynamics of $x$. Since neither $y$ nor $z$ appear in
the Lagrangian explicitly, we can instead integrate over $\dot y$ and $\dot z$
at the expense of constraining $\frac{1}{2\pi}\int dy$ and $\frac{1}{2\pi}
\int dz$ to be integers. This can be achieved with the discrete delta
functions
\begin{equation}
\int \mathcal Dy\, \mathcal Dz\, \to \sum_{k\in\ZZ}\sum_{m\in\ZZ} \int \mathcal D\dot y\, \mathcal D\dot z\, e^{ik\int dt\, \dot y}\, e^{im\int dt\, \dot z}\,.
\end{equation}
Integrating out $\dot y$ and $\dot z$, we find
\begin{align}
Z =\sum_{k\in\ZZ}\sum_{m\in\ZZ} \int \mathcal Dx\,\mathcal Dz\,\mathcal D\xi\,  e^{-\int dt\, \frac{1}{2g}\dot x^2+ V(x)+ \frac{g}{8\pi^2} \left(x + \xi - 2\pi k\right)^2-i\dot z\left(\frac{p}{2\pi}\xi+m\right)} \\
\sim \sum_{m=0}^{p-1}\sum_{k\in\ZZ}\int dt\, \mathcal Dx\, e^{-\int\frac{1}{2g}\dot x^2+ V(x)+ \frac{g}{8\pi^2} \left(x -\frac{2\pi m}{p}-2\pi k \right)^2} \equiv \sum_{m=0}^{p-1}Z_m.
\label{eq:Zqm} 
\end{align}
Upon integrating out the auxiliary fields the partition function becomes highly
non-local, decomposing into a sum of partition functions $Z_m$ labelled by the
corresponding value of the Lagrange multiplier $\chi = 2\pi m/p$.  The $p$
different terms on the right-hand side of \eqref{eq:Zqm} are the universes we
advertised earlier.   

On the other hand, if we restrict our attention to
constant values of $x$ and take the Euclidean time direction to be large,
then the effect of integrating out $y,z,\chi$ is to produce a local effective
potential, 
\begin{equation}
V_{\text{eff}}(x) = V(x) + \min_{k\in\ZZ}\frac{g}{8\pi^2} \left(x -\frac{2\pi k}{p} \right)^2\, ,\label{eq:Vqm} 
\end{equation}
with degenerate minima related by $(\ZZ_p)_x$. On the one hand, these $p$ vacua
appear to have matched the `t Hooft anomaly between $(\ZZ_p)_x$ and $(\ZZ_p)_z$.
At the same time, the naive quantum-mechanical expectation is that the true
ground state of the system is described by the $(\ZZ_p)_x$-symmetric linear
combination of wave functions localized at each minimum. Put differently, one
might expect instantons to lift the $p$-fold degeneracy by a non-perturbative
amount. However, there are no instanton configurations connecting the minima of
the potential \eqref{eq:Vqm}. To consider dynamical configurations we must pass
to the non-local expression \eqref{eq:Zqm}, where it becomes obvious that
`nearest-neighbor instantons' are not sensible field configurations because
they would have to connect different terms in the path integral decomposition
(i.e., universes). 
 
An immediate corollary is that the $(\ZZ_p)_x$ symmetry is spontaneously broken, contrary
to the usual lore prohibiting such a conclusion in quantum mechanics. To see
this, let us evaluate the expectation value $\langle e^{ix}\rangle$ with a small
$(\ZZ_p)_x$-breaking perturbation added to the action, and with time having finite extent $T$, 
\begin{align}
\langle e^{ix}\rangle &=\frac{1}{Z} \sum_{m=0}^{p-1}\sum_{k\in\ZZ}\int \mathcal Dx\,e^{ix}\, e^{-\int dt\, \frac{1}{2g}\dot x^2+ V(x)+ \frac{g}{8\pi^2} \left(x -\frac{2\pi m}{p}-2\pi k \right)^2-\epsilon \cos(x)} \\
&= \frac{\sum_{m=0}^{p-1}\langle e^{ix}\rangle_m \, Z_m(\epsilon)}{\sum_{m=0}^{p-1} Z_m(\epsilon)} = \frac{\sum_{m=0}^{p-1} \langle e^{ix}\rangle_m \, e^{-T\mathcal F_m(\epsilon)}}{\sum_{m=0}^{p-1} e^{-T\mathcal F_m(\epsilon)}}, \label{eq:expectationvalue} 
\end{align}
where $\langle e^{ix}\rangle_m$ is the expectation value in the $m$'th universe with free energy $\mathcal F_m(\epsilon)$. When $\epsilon >0$, the free energy $\mathcal F_0(\epsilon) < \mathcal F_i(\epsilon)$ for $i>0$. Taking the limit $T\to\infty$, only a single term in \eqref{eq:expectationvalue} survives,
\begin{equation}
\lim_{T\to\infty} \langle e^{ix}\rangle = \langle e^{ix}\rangle_0. 
\end{equation}
Finally, taking the symmetry-breaking parameter to zero, we find
\begin{equation}
\lim_{\epsilon\to0}\lim_{T\to\infty} \langle e^{ix}\rangle =\lim_{\epsilon\to0} \langle e^{ix}\rangle_0 = 1,
\end{equation}
so that the $(\ZZ_p)_x$ symmetry is spontaneously broken. Note that in coming to this conclusion, the exact decomposition of the path integral into universes was crucial--without it, finite-action instantons would proliferate even in the $T\to\infty$ limit, leading to a unique symmetric ground state. Finally, we note that the $p$ degenerate ground states match the 't Hooft anomaly
between the two $\ZZ_p$ symmetries. 

We can also see how these degenerate ground states appear from a Hamiltonian
perspective. In Minkowski space, the Hamiltonian resulting from the Lagrangian
\eqref{eq:qmexample} is
\begin{equation}
\hat H = \frac{g}{2}\left[\hat p_x^2 + \left(\hat p_y+\frac{\hat p_z}{p}-\frac{\hat x}{2\pi}\right)^2\right] +
V(\hat x,\hat y, \hat z) \,. \label{eq:Hqm} 
\end{equation}
In the Hamiltonian formalism the $(\ZZ_p)_x$ and $(\ZZ_p)_z$ shift symmetries
are generated by the operators $\hat U_x = e^{i\frac{2\pi}{p}\hat p_x}e^{-i\hat
z}$ and $\hat U_z = e^{i\frac{2\pi}{p}\hat p_z}$, respectively. Let $V(\hat x,\hat y, \hat z)=0$ for
simplicity. Then the above Hamiltonian is equivalent to a Landau problem on the
torus with $p$ units of magnetic flux~\cite{Komargodski:2017dmc}. The system has
$p$ degenerate ground states $|m\rangle$ on the torus identified by their
eigenvalues $\hat U_z|m\rangle  = e^{2\pi im/p}\,|m\rangle$ under $(\ZZ_p)_z$.
The wave functions are
\begin{equation}
\Psi_m(\vec{x})= \langle \vec{x}|m\rangle =\frac{1}{2\pi}\left(\frac{1}{2\pi^2}\right)^{1/4} \sum_{k\in\ZZ}\, e^{iky} e^{im z} e^{-\frac{1}{4\pi}\left(x- 2\pi k -\frac{2\pi m}{p}\right)^2}\, .
\end{equation}
The $p$ ground states have energy $E = g/4\pi$ and are permuted by $(\ZZ_p)_x$
as $\hat U_x|m\rangle = |m-1\rangle$. The symmetry operators therefore satisfy
the relation $\hat U_x \hat U_z = e^{2\pi i/p} \, \hat U_z \hat U_x$
corresponding to a central extension of $\ZZ_p \times\ZZ_p$ to the
Weyl-Heisenberg group. This is the manifestation of the mixed anomaly. 

\begin{figure}[h!] 
\centering
\subfigure[$\mu=0,\epsilon=0$]{\includegraphics[width=0.32\textwidth]{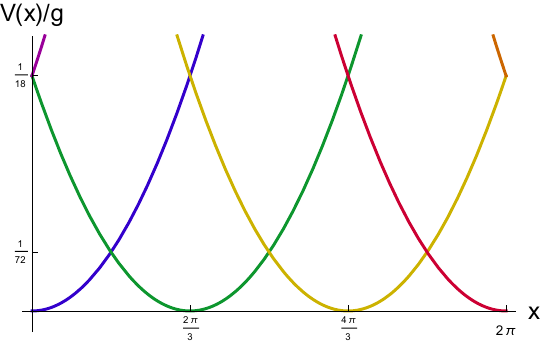}} 
\subfigure[$\mu=0,\epsilon\not=0$]{\includegraphics[width=0.32\textwidth]{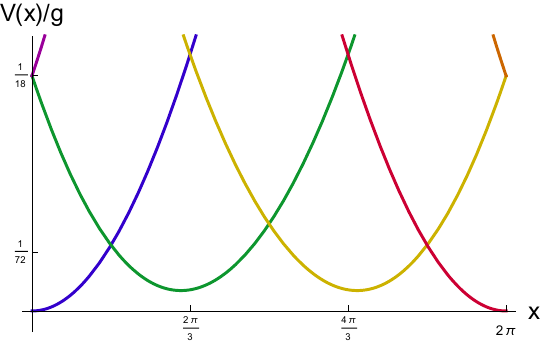}\label{fig:qm2}}
\subfigure[$\mu\not=0,\epsilon=0$]{\includegraphics[width=0.32\textwidth]{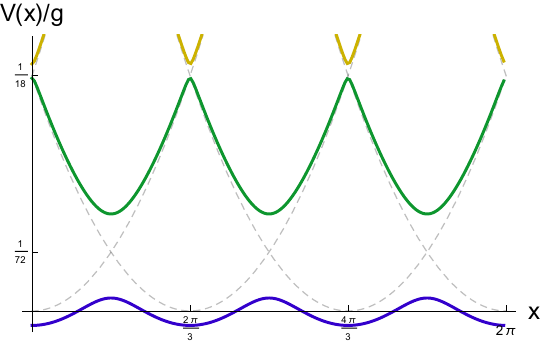}\label{fig:qm3}}
\caption{Effective potential $V_{\text{eff}}(x)$ for $p=3$, for different realizations of the $(\ZZ_p)_x$ and $(\ZZ_p)_z$ symmetries. The potential on the right is obtained by numerically diagonalizing the Hamiltonian \eqref{eq:Hqm2} with $\mu = .02\,  g$. The curvature of the potential near $x=0$ is roughly $m_{\text{eff}} \approx .02 \, g < \sqrt{p}\, \mu$, so the Born-Oppenheimer approximation is justified. } 
\label{fig:qm} 
\end{figure}

It is important to note that the existence of $p$ universes is protected by the
$(\ZZ_p)_z$ symmetry alone, and does not rely on the existence of the
$(\ZZ_p)_x$ symmetry nor the 't Hooft anomaly between them. To see this, we can
again add the $(\ZZ_p)_x$-breaking term $\epsilon\cos(x)$ to the Lagrangian. As
shown in Figure~\ref{fig:qm2}, such a term lifts the degeneracy between
universes, but the universes remain. Despite the fact that we can make the
difference in energy densities between neighboring vacua large by increasing
$\epsilon$, if we prepare the system in one of the false vacuum states it cannot
tunnel into the true vacuum. Hence, the false vacua are eternal. 

On the other hand, we can leave $(\ZZ_p)_x$ intact and break the $(\ZZ_p)_z$
symmetry explicitly by (for example) including a $V(\hat x,\hat y, \hat z) =
\cos(\hat z)$ potential,
\begin{equation}
\hat H = \frac{g}{2}\left[\hat p_x^2 + \left(\hat p_y+\frac{\hat p_z}{p}-\frac{\hat x}{2\pi}\right)^2\right] +\mu\cos (\hat z)\,. \label{eq:Hqm2} 
\end{equation}
This leaves a single universe with degenerate vacua related by the remaining
$(\ZZ_p)_x$ symmetry. Since these vacua now exist in the same universe,
finite-action tunneling events will proliferate and give rise to a unique
symmetric vacuum. In fact, the ground state remains unique even as we take $\mu
\to 0$. On the other hand, we know that when $\mu$ is strictly vanishing there
are $p$ degenerate ground states.  

To see this explicitly we can treat the variable $x$ within the Born-Oppenheimer
approximation. Integrating out the `fast' mode $z$ gives an effective
potential for $x$ which is the analog of Eq.~\eqref{eq:Vqm} when the $(\ZZ_p)_z$
symmetry is broken. The Born-Oppenheimer approximation is justified as long as
the curvature of the induced effective potential is smaller than $\sqrt{p}\,
\mu$, which can easily be arranged. Though the $(\ZZ_p)_z$-breaking potential is
non-linear, this process can be done numerically. As shown in
Figure~\ref{fig:qm3}, the level-crossings are avoided and the spectrum organizes
itself into bands. Within the lowest band, instantons proliferate and give rise
to a unique ground state. We can even compute the instanton action for
nearest-neighbor tunneling in the limit $\mu \to0$. In this limit the effective
potential reduces to~\eqref{eq:Vqm}. When $V(x) =0$, the instanton action is
$S_I = \frac{\pi}{2p^2}$, which clearly shows no singularity as $\mu\to0$. Again, this signals a non-uniformity in the symmetry-restoring limit: we know from our above arguments that $\mu = 0$ is a distinguished point in parameter space where tunneling is completely suppressed, i.e. where $S_I$ is infinite. 

\section{Universes in the massless Schwinger model and its deformations}
\label{sec:2d} 

We now move on to examples of universes and eternal false vacua in quantum field
theory. It is crucial to note that while in quantum mechanics an `eternal false
vacuum' is a trivial concept, it becomes nontrivial in relativistic QFT, where
vacua should be Poincar\'e-invariant. For instance, while an ordinary 0-form
global symmetry may ensure that certain single-particle states are stable, such
states are not eternal false vacua as they are not Poincar\'e-invariant. Indeed,
eternal false vacua can only exist due to $(d-1)$-form symmetries, which are
higher-form symmetries in $d \ge 2$. 

In this section we analyze variations of the Schwinger model: two-dimensional
$U(1)$ gauge theory with a charged Dirac fermion. We begin by taking the fermion
to be massless, with charge $p \in \ZZ$, with $|p|>1$. Many aspects of the
Schwinger model with non-minimal charge have been explored in recent
works~\cite{Anber:2018jdf,Anber:2018xek,Misumi:2019dwq,Armoni:2018bga,Komargodski:2017dmc,Komargodski:2020mxz},
where the mixed anomaly between the 0-form chiral and 1-form center symmetries
plays a central role. Reference~\cite{Anber:2018xek} identified the $p$ vacuum
branches of the theory as belonging to different universes (this was also
discussed more recently in \cite{Komargodski:2020mxz} using the language of
universes), and explained how (de)confinement can be understood in terms of the
energy densities in each universe. While our discussion in this section contains
some overlap with Refs.~\cite{Anber:2018xek,Komargodski:2020mxz}, our main goal is to provide a
self-contained discussion of how universes arise in the model, and in particular
how eternal false vacua appear when the fermion is given a finite mass. We give
parametric estimates of the lifetimes of false vacua when both chiral and center
symmetry-breaking deformations are included. A discussion of false vacua similar
in spirit to our analysis appears in Ref.~\cite{Lawrence:2012ua} which,
motivated by axion monodromy models of inflation, examines two-dimensional
theories featuring multi-branched potentials with metastable vacua. 

In Section~\ref{sec:modified_sum}, we show how the charge-$p$ Schwinger model
can be understood as a charge-$1$ Schwinger model with a modified instanton sum.
This perspective allows us to generalize our discussion of universes and eternal
false vacua in two-dimensions to higher dimensions. We discuss the symmetries
and $\theta$-dependence of the modified model in detail. 

\subsection{Charge-$p$ Schwinger model}
\label{sec:chargep_schwinger} 

The Euclidean Lagrangian of the massless charge-$p$ Schwinger model is
\begin{equation}
\mathcal L = \frac{1}{4e^2}f_{\mu\nu}f^{\mu\nu} 
+\frac{i\theta}{2\pi}\epsilon^{\mu\nu}\partial_\mu a_\nu 
+ \bar\psi\gamma^\mu(\partial_\mu-ip\, a_\mu)\psi\, .
\end{equation}
where $f_{\mu\nu} = \partial_{\mu} a_{\nu} - \partial_{\nu} a_{\mu}$.  We assume
that the theory is defined on a closed 2-manifold so that the topological
charges are quantized as
\begin{equation}
\int \epsilon^{\mu\nu}\partial_\mu a_\nu =  \int da 
\in 2\pi\ZZ .
\end{equation}
The theory has a $\ZZ_{2p}^{(0)}$ 0-form chiral symmetry, but the $\ZZ_2$
subgroup generated by $(-1)^F$ is actually non-chiral and lies within the $U(1)$
gauge group. We will focus on the faithfully acting $\ZZ_p^{(0)} =
\ZZ_{2p}^{(0)}/\mathbb{Z}_2$ chiral symmetry below. The theory also has a 1-form
$\ZZ_p^{(1)}$ center symmetry. The bosonized form of the massless charge-$p$
Schwinger model is~\cite{PhysRevD.11.2088,Coleman:1976uz}
\begin{equation}
\mathcal L  =\frac{1}{2e^2}|da|^2 + \frac{1}{8\pi}|d\varphi|^2 
+\frac{ip}{2\pi}\varphi\wedge da + \frac{i\theta}{2\pi} da, \label{eq:bosonized_schwinger} 
\end{equation}
where we have switched to form notation, so that $a = a_{\mu}dx^{\mu}$ and used
the shorthand $|\omega|^2 = \omega\wedge\star\omega$. Here $\varphi$ is a
$2\pi$-periodic scalar and the discrete chiral symmetry becomes the shift
symmetry $\varphi \to \varphi + 2\pi/p$. The 1-form $\ZZ_p^{(1)}$ symmetry acts
by shifting $a \to a + \lambda$ by a flat connection with $\int \lambda = 2\pi
\ZZ/p$. The $\ZZ_p^{(0)}$ and $\ZZ_p^{(1)}$ symmetries have a mixed
anomaly~\cite{Anber:2018jdf,Misumi:2019dwq,Armoni:2018bga}, which can be seen by
turning on an appropriate background 2-form gauge field for the 1-form symmetry.


An important point is that the symmetry operator of the
$1$-form symmetry is a local operator, which up to a normalization constant is
given by $U_m(x) = e^{i \frac{2\pi m}{p} \left( -\frac{i }{e^2}\star da +
\frac{p}{2\pi} \varphi\right)}$.  Indeed, the equation of motion for the gauge field can be written as 
\begin{align}
    d\left(  -\frac{i }{e^2}\star da +
    \frac{p}{2\pi} \varphi \right) = 0 \, .
\end{align}
The 0-form expression in parentheses is therefore a conserved quantity, and this
can be used to show that $U_m(x)$ is a well-defined local topological operator
so long as $m\in \mathbb{Z}$. The symmetry operator satisfies 
\begin{equation}
\langle U_m(x) e^{i\int_C a}\rangle = e^{\frac{2\pi i m}{p} \ell(C,x)}\langle e^{i \int_C a}\rangle,
\end{equation}
where the linking number $\ell(C,x) = 0$ or $1$ depending on if the point $x$ is
inside the closed contour $C$, which we take to be oriented and without
self-intersections. So long as the 1-form symmetry is not explicitly broken,
correlation functions of $U_m(x)$ should be topological --- which in the case of
local operators, means independent of the position $x$. We will see that this
enforces strict constraints on the possible dynamical domain wall configurations
in which $\varphi$ approaches distinct values at asymptotic infinity. In
particular, domain wall configurations where $U_m(x)$ depends on spacetime
necessarily have infinite tension. While one can consider such domain walls  as
probes of the theory, they are not \emph{dynamical}
excitations~\cite{Komargodski:2020mxz}.

In the deep IR limit $e^2 \to \infty$, integrating out $a$ locally sets
$d\varphi = 0$, while the sum over topologically non-trivial sectors (global
fluxes $\int da \in 2\pi\ZZ$) results in the constraint $\varphi = 2\pi \ell/p$,
with $\ell \in \ZZ$. The field $\varphi$ is frozen to one of $p$ values, and
there are no domain walls which interpolate between them. In this limit the
distinct values $\ell \in \{0,\ldots,p-1\}$ label distinct universes. Even
away from the extreme IR limit the theory \eqref{eq:bosonized_schwinger} is
Gaussian and we can integrate out the gauge field by treating $F=da$ as the
fundamental degree of freedom. Imposing $U(1)$ flux quantization (read: the
Bianchi identity) as a delta function constraint, 
\begin{equation}
\sum_{k \in\ZZ} e^{i k \int F} = \sum_{\nu \in \ZZ} \delta\left( \nu - \frac{1}{2\pi}\int F\right),
\end{equation} 
the path integral can be expressed as
\begin{equation}
Z(\theta) = \int\mathcal D\varphi \sum_{k\in\ZZ}\, \exp\left\{-\int \frac{1}{8\pi}|d\varphi|^2 + \frac{1}{2}\left(\frac{e\, p}{2\pi}\right)^2\left(\varphi + \frac{\theta}{p}-\frac{2\pi k}{p}\right)^2 \right\}. \label{eq:decomp} 
\end{equation}
There is no longer a local effective Lagrangian for $\varphi$: the partition
function decomposes into a sum of distinct path integrals labelled by the
integer $k$.\footnote{If we decompose $k = p\, n + \ell$, where $n \in \ZZ$ and
$\ell \in \{0,\ldots,p-1\}$, then $\ell$ can be interpreted as a discrete theta
parameter of the charge-$p$ Schwinger model with the $\ZZ_p^{(1)}$ 1-form
symmetry gauged. At a technical level, minimizing over $n$ ensures that the
potential is $2\pi$-periodic with respect to $\varphi$ while the minimization
over $\ell$ ensures the potential is $2\pi$-periodic with respect to $\theta$. }
The $p$ inequivalent terms in the above decomposition are
distinct universes, labelled by the expectation value of the $1$-form symmetry
operator $\langle U_1(x)\rangle = e^{2\pi i k/p}$. 

In the infinite volume limit, the potential for the zero mode of $\varphi$, which we denote
$\varphi_0$, can be written in a manifestly local way as
\begin{equation}
V(\varphi_0) = \min_{k\in\ZZ} \frac{1}{2}\left(\frac{e\, p}{2\pi}\right)^2\left(\varphi_0 + \frac{\theta}{p}-\frac{2\pi k}{p}\right)^2. \label{eq:zeromode} 
\end{equation}
The potential in Eq.~\eqref{eq:zeromode} has $p$ degenerate minima corresponding
to $\varphi_0 = (2\pi k -\theta)/p$. These minima are related by the
$\ZZ_p^{(0)}$ discrete shift symmetry, which is spontaneously broken. We emphasize that each chiral vacuum lies in a distinct universe. Hence, dynamical domain walls between neighboring vacua do not exist as
they violate the `topological' property of the operator $U_m(x)$ (which
is protected by the absence of charge-$1$ matter and the resulting 1-form
$\ZZ_p^{(1)}$ symmetry). As a result, these states do not mix even when
spacetime is compact: they belong to different universes. At the technical
level, domain walls cannot exist because each $\ZZ_p^{(0)}$-breaking vacuum lies
within a distinct branch of the effective potential.

\begin{figure}[h!] 
\centering
\subfigure[$p=1$]{\includegraphics[width=0.46\textwidth]{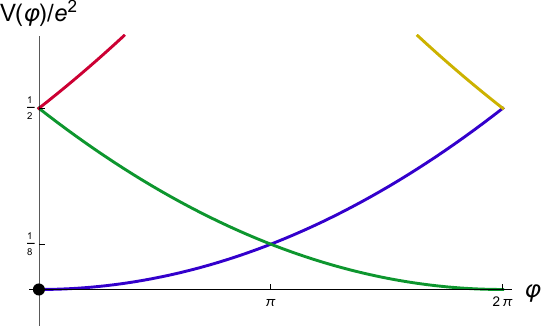}} \hspace{.5cm}
\subfigure[$p=3$]{\includegraphics[width=0.46\textwidth]{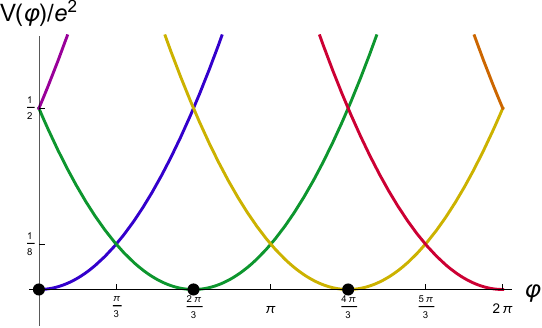}}
\caption{Multi-branched structure of the effective potential for massless Schwinger models at $\theta =0$ with charge $p=1$ and $p=3$. For $p=1$, all branches correspond to the same universe. For $p=3$, there are three distinct universes. } 
\label{fig:m=0} 
\end{figure}

At the risk of being pedantic, let us explain how spontaneous symmetry breaking
occurs in the presence of universes. To conclude that a (0-form) symmetry is
spontaneously broken one works in a large but finite spatial volume $V$ and
introduces a symmetry-breaking perturbation at some scale $m$ which is small
compared to the intrinsic scales (inverse correlation length) of the problem.
Normally, we say that a symmetry is spontaneously broken when
\begin{equation}
\lim_{m\to 0}\lim_{V \to \infty} \langle \mathcal O \rangle \not=0 \label{eq:ssb} 
\end{equation}
for some local operator $\mathcal O$ which transforms under the symmetry.
Importantly, the limits above do not commute, and it is always true that $\lim_{V
\to \infty}\lim_{m\to 0} \langle \mathcal O \rangle =0$. 

In the context of the Schwinger model, $m$ is a small chiral
symmetry-breaking mass for the fermion, with the $\theta$ parameter absorbed
into the phase of the fermion mass by a chiral rotation, $\arg(m) = \theta/p$.
Using Eq.~\eqref{eq:decomp} we can write the expectation value of $e^{i\varphi}
\sim \bar{\psi}_L \psi_R$ as
\begin{align}
\langle e^{i\varphi} \rangle_{m=0} = \lim_{\substack{m\to0, \\ \arg m \textrm{ fixed}}}
\lim_{L\to\infty} 
\frac{\sum_{k=0}^{p-1} \int \mathcal D\varphi \, e^{i\varphi} \, e^{-S_k(m)}}
{\sum_{k=0}^{p-1} \int \mathcal D\varphi \, e^{-S_k(m)}} \\
= \lim_{\substack{m\to0, \\ \arg m \textrm{ fixed}}}
\lim_{L\to\infty} 
\frac{\sum_{k=0}^{p-1}  \langle e^{i\varphi}\rangle_k\, Z_k(m)}
{\sum_{k=0}^{p-1} Z_k(m) }, \nonumber
\end{align}
where $\langle e^{i\varphi}\rangle_k$ is the expectation value in the universe
with action $S_k(m)$ and $L$ is the spatial volume.  It is useful to trade $Z_k$
for the free energy $\mathcal{F}_k$ via $Z_k(m) = e^{-LT\mathcal F_k(m)}$, where
$LT$ is the spacetime volume.  For any given fixed $|m|$ and a generic value
$\arg(m)$ the set $\{\mathcal{F}_k(m)\}$ has a unique minimum $\mathcal{F}_{\rm
k_{\rm min}}(m)$, but of course $k_{\rm min}$ depends on $\arg m$.\footnote{When
$\theta = \pi$ the minimum of $\{\mathcal{F}_k(m)\}$ becomes doubly-degenerate, corresponding to the spontaneous breaking of charge conjugation.
We will not consider this special case here.} As a result, in the infinite
volume limit only a single term in the sum contributes to the expectation value,
\begin{align}
\langle e^{i\varphi} \rangle_{m=0} = \lim_{\substack{m\to0, \\ \arg m \textrm{ fixed}}}\lim_{L\to\infty}
\frac{\sum_{k=0}^{p-1} \langle e^{i\varphi}\rangle_k \, e^{-LT\mathcal F_k(m)}}
{\sum_{k=0}^{p-1} e^{-LT\mathcal F_k(m)}} 
= \lim_{\substack{m\to0, \\ \arg m \textrm{ fixed}}}\langle e^{i\varphi}\rangle_0 = e^{2\pi i n/p} \, ,
\end{align}
where the value of the integer $n$ depends on $\arg m = \theta/p$.  If instead
we had taken $m\to0$ before $L\to\infty$, each universe would contribute equally
to the expectation value, and we would erroneously conclude that $\langle e^{i
\varphi}\rangle = \sum_{k=0}^{p-1}e^{2\pi i k/p} = 0$. 

We have seen that the $\ZZ_p^{(0)}$ chiral symmetry is spontaneously broken in
the massless Schwinger model. When the fermion is massless the $\ZZ_p^{(1)}$
symmetry is also spontaneously broken. This is easily seen by noticing that the
insertion of a charge-$q$ Wilson loop in the path integral shifts the theta
angle by $\theta \to \theta - 2\pi q$ inside the loop. Hence, inside the loop,
the partition function is that of the universe with an unshifted $\theta$ but
with $k\to k+q$ relative to the outside. Since the universes are degenerate, the
difference in energy density inside and outside the Wilson loop vanishes. As a
result there is no area-law contribution to the Wilson loop expectation value,
and $\ZZ_p^{(1)}$ is spontaneously broken. When the fermion mass $m\not=0$ the
theory becomes confining, with string tensions scaling like $m\, e$ or $m^2$
depending on whether $|m| \ll e$ or $|m| \gg e$~\cite{Komargodski:2020mxz}. Of
course, when we add charge-$1$ matter to the theory the 1-form symmetry is
explicitly broken. In the following, we examine both symmetry-breaking
perturbations of the massless Schwinger model in more detail.

\begin{figure}[h!]  
\centering
\subfigure[$m=0$]{\includegraphics[width=0.46\textwidth]{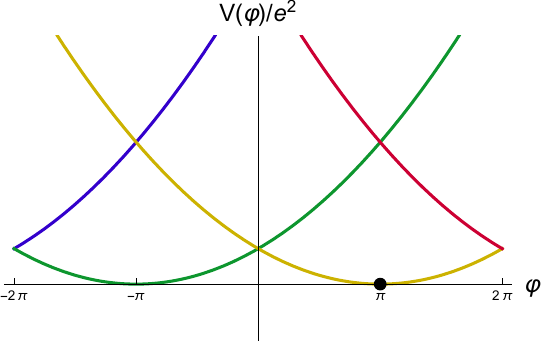}}  \hspace{.5cm}
\subfigure[$m>0$]{\includegraphics[width=0.46\textwidth]{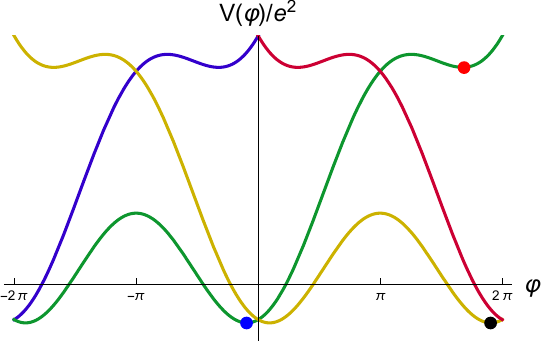}}
\caption{Multi-branched structure of the effective potential for $p=1,\theta =\pi$. 
The metastable vacuum (red dot) in the $k=0$ branch unstable. 
It decays to the minimum marked by the blue dot (on the same $k=0$ branch). 
By a shift of $2\pi$, this is equivalent to one of the true global minima, 
marked by the black dot.  } 
\label{fig:p=1}
\end{figure}

Let us explicitly introduce a mass for the fermion, so that the bosonized
potential is
\begin{equation}
V(\varphi_0) = \min_{k\in\ZZ} \frac{1}{2}\left(\frac{e\, p}{2\pi}\right)^2\left(\varphi_0 + \frac{\theta}{p}-\frac{2\pi k}{p}\right)^2 -  m\mu \cos\varphi_0,
\end{equation}
where $\mu$ is some scale determined by matching to the original fermionic
theory. The mass term breaks the $\ZZ_p^{(0)}$ symmetry completely, and for
generic values of $m,\theta$ there is a unique vacuum.  The exception is at
$\theta =\pi$, where two ground states emerge as one increases the mass. As an
example, consider the $p=1$ case at $\theta =\pi$,  
\begin{equation}
V(\varphi_0) = \min_{k\in\ZZ} \frac{1}{2}\left(\frac{e}{2\pi}\right)^2\left(\varphi_0 + \pi-2\pi k\right)^2 - m\mu \cos\varphi_0. 
\end{equation}
For $m\mu \ll e^2$ the potential has a single minimum in the
fundamental domain $\varphi_0\in(0,2\pi)$ at $\varphi_0 = \pi$. This
minimum lies on the branch labelled by $k=1$. As we increase $m$ past
a critical value, the potential
develops two ground states related by charge conjugation. As we
increase $m$ further, there is another critical value where local minima appear in the $k=0$ and $k=2$
branches, see Figure~\ref{fig:p=1}. These metastable vacua can decay
to one of the true vacua via a tunneling process. All of the branches
can be connected via similar tunneling processes--this is possible
because they lie within the same universe. 

\begin{figure}[h!] 
\centering
\subfigure[$m=0$]{\includegraphics[width=0.46\textwidth]{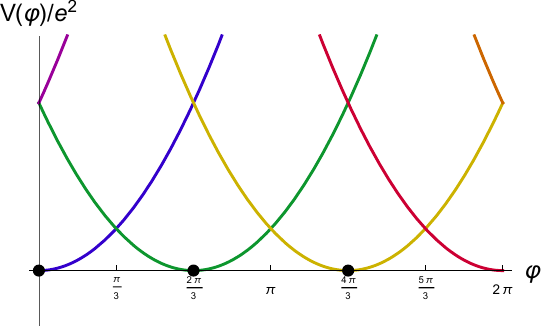} \label{fig:p=3a}} \hspace{0.5cm}
\subfigure[$m>0$]{\includegraphics[width=0.46\textwidth]{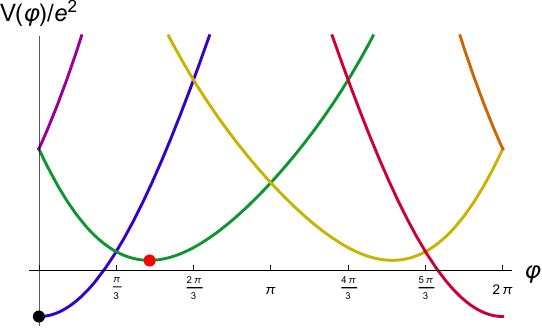} \label{fig:p=3b}}
\caption{Multi-branched structure of the effective potential for $p=3,\theta =0$.} 
\label{fig:p=3} 
\end{figure}

Let us contrast this to what happens when $p>1$. In Figure~\ref{fig:p=3} we compare
the $p=3$ effective potential at $\theta =0$ for zero and nonzero
mass. When $m=0$ there are three degenerate ground states which
spontaneously break $\ZZ_3^{(0)}$. Each degenerate vacuum lies in a
distinct universe. As soon as the fermion mass is nonzero the
degeneracy is lifted and there is a unique vacuum which lies in the
branch labelled by $k=0$ mod $3$. False vacua in different branches
cannot decay to the true vacuum and are infinitely long-lived. 

\begin{figure}[h!] 
\centering
\includegraphics[width=0.75\textwidth]{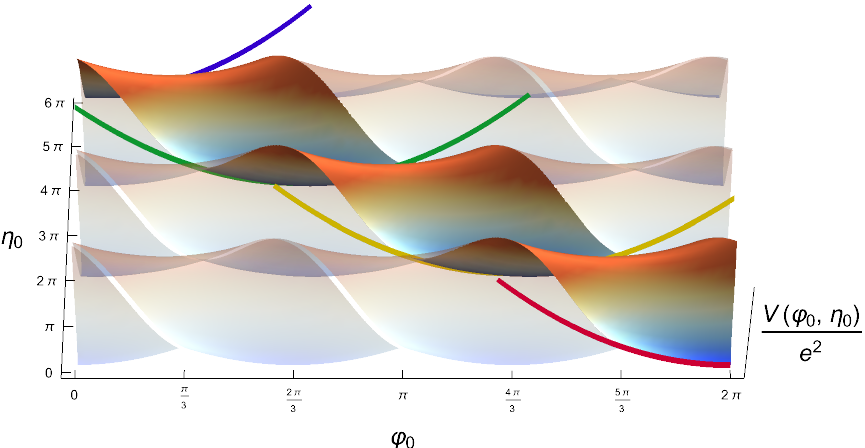}
\caption{Effective potential for bosonized scalars $\varphi,\eta$ with $p=3, m
=0, M>0$. The four curves coincide with the four branches in
Figure~\ref{fig:p=3a}. } 
\label{fig:p=31} 
\end{figure}

The infinite lifetimes of false vacua in the above examples with $p>1$ were
protected by the 1-form $\ZZ_p^{(1)}$ symmetry. This symmetry is explicitly
broken when we introduce matter with minimal charge.  To this end, let us
introduce a heavy charge-$1$ fermion $\zeta$ with mass $M$, with $M^2\gg e^2$.
When $p$ is even, the $\zeta$ fermion number $(-1)^{F_\zeta}$ coincides with a
gauge transformation, while when $p$ is odd, the overall fermion number $(-1)^F$
lies within the gauge group. Applying abelian bosonization to $\psi
\leftrightarrow \varphi$ and $\zeta\leftrightarrow \eta$~\cite{Coleman:1976uz},
we obtain a description of the theory with the overall fermion number gauged, 
\begin{equation}
\mathcal L = \frac{1}{2e^2}|da|^2+\frac{1}{8\pi}|d\varphi|^2 
+ \frac{1}{8\pi}|d\eta|^2 + 
\frac{ip}{2\pi}\left(\varphi+\frac{\eta+\theta}{p}\right)\wedge da - c\, M^2\cos\eta, \label{eq:bosonizedp1} 
\end{equation}
where $c$ is an $\mathcal O(1)$ constant. Again, we can integrate out the gauge
field to obtain the effective potential for the zero-modes of $\varphi$ and
$\eta$. This is shown in Figure~\ref{fig:p=31} for $p=3$. The bosonized
charge-$1$ fermion provides an extra direction in field space, and the result is
that the branches in Figure~\ref{fig:p=3a} become smoothly connected: they are
all part of the same universe. 

\begin{figure}[h!] 
\centering
\includegraphics[width=0.75\textwidth]{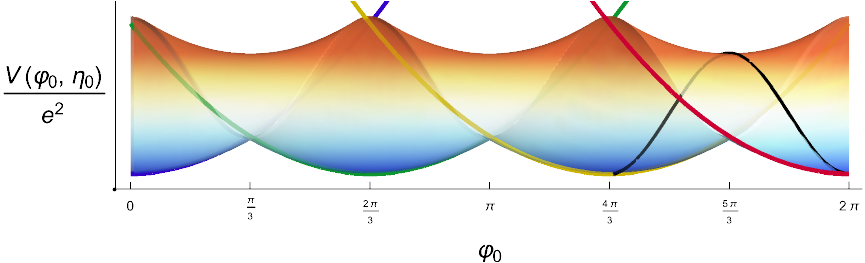}
\caption{Domain wall configuration between two chiral symmetry breaking vacua
for $p=3$, $\theta=0$. Rather than pass through the cusp, the domain wall
trajectory must go above the barrier, whose height scales as $M^2$.  The curves coincide with the four branches in
Figure~\ref{fig:p=3a}.} 
\label{fig:domainwall} 
\end{figure}

Unlike when the $\ZZ_p^{(0)}$ symmetry is exact, now there are finite tension
domain wall configurations connecting discrete chiral vacua. On the domain wall
$\eta$ interpolates between $0$ and $2\pi$ while $\varphi$ interpolates between
$2\pi/p$ and $0$. As an example, the trajectory for $p=3$ connecting the $(\varphi,\eta) =
(2\pi,0) \to (4\pi/3,2\pi)\sim(4\pi/3,0)$ vacua is shown in
Figure~\ref{fig:domainwall}. The domain wall connecting neighboring chiral vacua
follows a straight trajectory in field space parameterized by $\varphi(\tau) =
\frac{2\pi k}{p}-\alpha(\tau), \eta = p\, \alpha(\tau)$. In terms of $\alpha$,
the domain wall is simply a Sine-Gordon kink 
\begin{equation}
\alpha(\tau) = \frac{4}{p}\arctan\left[ \exp\left(\sqrt{\frac{4\pi p\, c}{1+p^2}}\,M \tau\right)\right],
\end{equation}
with tension $T \sim M$ up to $\mathcal O(1)$ factors. 

As we increase $M$ the
height of the barrier between vacua increases in the smooth direction
in field space, while the height of the barrier in the cuspy direction
where only $\varphi$ varies stays fixed. When $M$ is large, even
though the domain wall passing through the large barrier seems more
costly than the domain wall passing through the cusp, the latter
configuration is still not admissible. When $M \to \infty$ there
are no finite-tension domain walls. Indeed, in the formal limit
$M\to\infty$ the unit charge particle decouples and we land on
the charge-$p$ model which has $p$ universes.

If we also introduce a mass $m$ for the charge-$p$ fermion the
chiral symmetry is broken completely. When $\theta =0$ there is a
single unique vacuum at $\varphi= \eta=0$. Unlike the theory without
charge-$1$ fields (see Figure~\ref{fig:p=3b}), the metastable vacua can
decay via the nucleation of bubbles of true vacuum.  
\begin{figure}[h!] 
\centering
\includegraphics[width=0.6\textwidth]{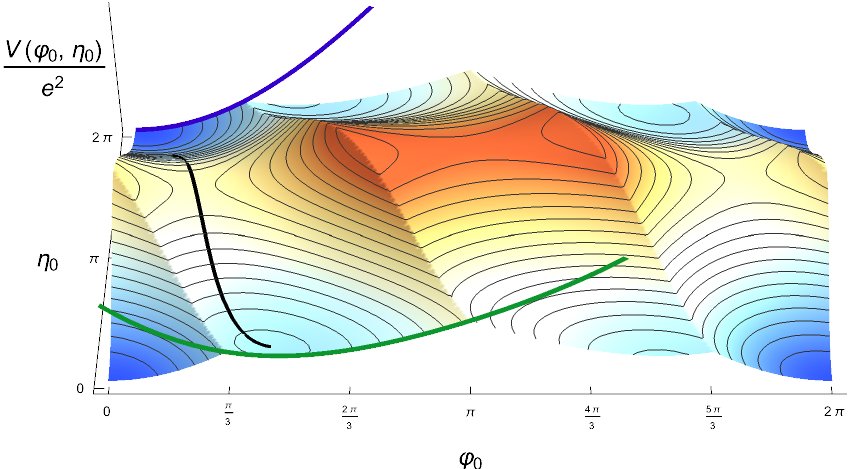}
\caption{Effective potential for $p=3$, $\theta=0$, $m>0$, and $M>0$. The bounce
solution renders the false vacuum unstable. The blue and green curves coincide with the
 branches in Figure~\ref{fig:p=3b}.} 
\label{fig:bounce} 
\end{figure}
Despite the fact that the false vacuum has an energy density scaling
with $m$ relative to the true vacuum, the decay rate can be made
arbitrarily small by increasing $M$. The decay rate of the false
vacuum scales as $\sim e^{-S_I}$ where $S_I$ is the action of a bounce
solution, which has the form
\begin{equation}
S_I = - \pi R^2 \, \varepsilon + 2\pi R\,  T,
\end{equation}
where $\varepsilon \sim m\mu$ is the difference in energy density of the two
minima and $T$ is the tension of the bounce solution, which to leading order in
a large $M$ expansion is given by the domain wall tension $T \sim M$. The size
of the bubble of true vacuum is $R= T/\epsilon$, so the action of the bounce
configuration $S_I= \pi  T^2/\epsilon$ scales as $M^2/(m\mu)$ for $M$ large.
When $m$ is small compared to $|e|$, $\mu \sim e$. Hence, the lifetime of the
false vacuum is exponentially enhanced by the ratio of a `UV' scale $M$ and the
low-energy scales $e$ and $m$,
\begin{equation}
\tau \sim e^{S_I} \sim e^{M^2/(e\, m)},
\end{equation}
leading to near-eternal false vacua when $M^2 \gg e\, m$.

\subsection{Charge-$1$ Schwinger model with a modified instanton sum}
\label{sec:modified_sum}

So far we have discussed the massless and massive charge-$p$ Schwinger models,
and their extensions to include charge-$1$ fermions. Here we provide a different
construction that leads to the same physics. As pointed out in Refs.~\cite{Pantev:2005rh,Seiberg:2010qd}, restricting the instanton number in the 2d $\mathbb{C}\mathbb{P}^{N-1}$ model to
be a multiple of $p$ gives rise to the same physics as the charge-$p$
$\mathbb{C}\mathbb{P}^{N-1}$ model. Indeed, this
correspondence applies to all $U(1)$ gauge theories in two dimensions. For $d>2$
the relation between modified instanton sums and non-minimal charge assignments
breaks down. However, theories with modified instanton sums share the features
of 2d abelian theories with charge-$p$ matter, namely the existence of universes
and false vacua. To set the stage for our treatment of 4d theories with modified
instanton sums, we explain the 2d correspondence in detail.

Consider a generic 2d $U(1)$ gauge theory with a charge-$1$ fermion (a nearly
identical discussion holds for bosonic matter), 
\begin{equation}
\mathcal{L} = \frac{1}{2e^2} |da|^2 
+ \bar{\psi} (\gamma^{\mu} \partial_{\mu}+ i a) \psi 
+ \frac{i\theta}{2\pi}da + \cdots, 
\end{equation}
where the ellipses denote other terms allowed by gauge invariance (we do not impose
chiral symmetry for the moment).  We now modify the theory so that only topological
charges which are multiples of $p$ contribute to the partition function~\cite{Seiberg:2010qd}. To this
end, let us introduce another $U(1)$ gauge field $\hat a$ and a $2\pi$-periodic
compact scalar Lagrange multiplier $\chi$, so that the Lagrangian is shifted
by 
\begin{equation}
\delta \mathcal{L} = \frac{i}{2\pi}\chi\wedge(da-p\, d\hat a) + \frac{i\hat{\theta}}{2\pi} d \hat{a} 
\label{eq:2d_modified_sum}
\end{equation}
where $\hat{\theta}$ is a new  $2\pi$-periodic theta parameter. Integrating out $\chi$ sets $da=p\,
d\hat a$, which means
\begin{equation}
\frac{1}{2\pi}\int da = \frac{p}{2\pi}\int d\hat a \in p \ZZ, \label{eq:2drestriction} 
\end{equation}
so that only topological charges (for $a$) which are multiples of $p$ contribute
to the path integral. As a result, $\theta$ has a reduced periodicity, $\theta
\sim \theta + 2\pi/p$. 

On the other hand, the equation of
motion for $\hat a$ sets $\chi =$ constant, while integrating out $\hat a$
and performing the sum over its topological sectors sets $\chi = 2\pi k/p$
for $k=0,\ldots,p-1$.  The path integral sums over $\ell$, so that
\begin{align}
Z_{\rm constrained}(\theta) = \sum_{k = 0}^{p-1} Z_{\rm original} (\theta+ 2\pi k/p), \label{eq:decomposition} 
\end{align}
which makes manifest the $2\pi/p$-periodicity of $\theta$. 

To see that the charge-$1$ theory with a modified instanton sum is equivalent to
the corresponding charge-$p$ theory with a standard instanton sum, we simply
solve the local constraint imposed by $\chi$. Substituting $da=p\, d\hat a$ into
the Lagrangian,
\begin{equation}
\mathcal{L} = \frac{p^2}{2e^2} |d\hat{a}|^2 + 
\bar{\psi} (\gamma^{\mu} \partial_{\mu}+ i p\, \hat{a}) \psi 
+ \frac{i(p \, \theta+\hat{\theta} )}{2\pi}d \hat{a}. 
\end{equation}
If we now rescale the gauge coupling $e = p\, \hat{e}$, and define
$\tilde{\theta} = p\, \theta+ \hat{\theta} $, this is just a charge-$p$ $U(1)$
gauge theory with $\theta$ parameter $\tilde{\theta}$ and no modification to the
instanton sum. The gauge coupling rescaling is harmless --- the only role of $e$
is to set the overall energy scale for the theory. 

All of the discussions above are about universes in 2d gauge Abelian gauge theory
coupled to charge-$p$ matter applies verbatim to the corresponding theories with
minimally-charged matter and modified instanton sums. Uplifting the former
construction to higher dimensions only gives rise to a $\ZZ_p$ 1-form symmetry, while
the latter gives rise to the $\ZZ_p$ $(d-1)$-form symmetry necessary for the
existence of universes and eternal false vacua. 

To make the connection to the charge-$p$ model more explicit, it is worth discussing the 0-form chiral symmetry of the
charge-$1$ Schwinger model with a modified instanton sector, whose bosonized
form is
\begin{align}
\mathcal L = \frac{1}{2e^2}|da|^2 + \frac{1}{8\pi}|d\varphi|^2 + \frac{i}{2\pi}\varphi\wedge da + \frac{i}{2\pi}\chi\wedge (da-p\, d\hat a). \label{eq:modifiedschwinger} 
\end{align}
Using the arguments in the previous paragraph, this theory is equivalent to the
charge-$p$ Schwinger model. One might ask what, in the modified instanton sum
description, is the analog of the discrete chiral symmetry $\ZZ_p^{(0)}$ of the
charge-$p$ Schwinger model. The fact that $e^{i\chi}$ becomes, after integrating
out $\hat a$, a discrete field valued in the $p$th roots of unity leads to a
mild violation of cluster decomposition also observed in the context of
`Gerby' conformal field theories~\cite{Hellerman:2006zs,Pantev:2005rh}.
Relatedly, the symmetry transformation properties of $e^{i\chi}$ are subtle and
should be treated carefully. 

In the bosonized variables, a first guess might be
to define a shift symmetry which takes $\varphi \to \varphi + 2\pi/p$ and does
not act on $\chi$.  For reference, let us compute the path integral over $\hat
a$ and $\chi$,
\begin{align}
\int \mathcal D \hat a\, \mathcal D\chi \, e^{-\frac{i}{2\pi}\int\varphi\wedge da-\frac{i}{2\pi}\int\chi\wedge(da-p\, d\hat a)}
 &= \int \mathcal D\hat f\, \mathcal D\chi \sum_{k\in\ZZ}\, e^{i\int\left(k+\frac{p}{2\pi}\chi\right) \hat f} e^{-\frac{i}{2\pi}\int\varphi\wedge da-\frac{i}{2\pi}\int\chi\wedge da} \\
&=\frac{2\pi}{p}\sum_{k\in\ZZ}\int \mathcal D\chi \,\delta\left(\chi+\frac{2\pi k}{p}\right) e^{-\frac{i}{2\pi}\int\varphi\wedge da-\frac{i}{2\pi}\int\chi\wedge da} \\
&= \frac{2\pi}{p}\sum_{k\in\ZZ} \, e^{\frac{ik}{p}\int da}\, e^{-\frac{i}{2\pi}\int\varphi\wedge da} \\
&= 2\pi \sum_{m\in\ZZ} \delta\left(\frac{1}{2\pi}\int da-p\, m \right) e^{-\frac{i}{2\pi}\int\varphi\wedge da}, \label{eq:Ztqft} 
\end{align}
where in the first equality we defined $\hat f = d\hat a$. As expected, the TQFT
sector in \eqref{eq:modifiedschwinger} simply enforces a constraint on the
topological charge of the gauge field $a$.

The shift $\varphi\to \varphi+2\pi/p$ does not leave the exponentiated action
$e^{-S}$ invariant. However, it does leave the partition function invariant,
since 
\begin{align}
\int \mathcal D \hat a\, \mathcal D\chi \, e^{-\frac{i}{2\pi}\int\varphi\wedge da-\frac{i}{2\pi}\int\chi\wedge(da-p\, dc)} \to 2\pi \sum_{m\in\ZZ} \delta\left(\frac{1}{2\pi}\int da-p\, m \right) e^{-\frac{i}{2\pi}\int\varphi\wedge da}\, e^{-\frac{i}{p}\int da}.
\end{align}
This is equivalent to \eqref{eq:Ztqft} because the delta function setting $\int
da \in 2\pi p\,  \ZZ$ ensures that the extra phase $e^{-\frac{i}{p}\int da}$ is
trivial. The invariance of the path integral is only manifest \emph{after}
integrating out both $\hat a$ and $\chi$. Now let us calculate the
expectation values of $\langle e^{i\varphi}\rangle$ and $\langle
e^{i\chi}\rangle$. Again, just from the path integral over $\hat a$ and
$\chi$, we have
\begin{align}
\int \mathcal D \hat a\, \mathcal D\chi \,e^{i\varphi}\,  e^{-\frac{i}{2\pi}\int\varphi\wedge da-\frac{i}{2\pi}\int\chi\wedge(da-p\, d\hat a)} = 2\pi \sum_{m\in\ZZ} \delta\left(\frac{1}{2\pi}\int da-p\, m \right)e^{i\varphi}\,  e^{-\frac{i}{2\pi}\int\varphi\wedge da}\, .
\end{align}
From this it is clear that under the shift  $\varphi \to \varphi+2\pi/p$, we
have 
\begin{align}
    \langle e^{i\varphi} \rangle 
    \to e^{2\pi i/p}\, \langle e^{i\varphi} \rangle
    \label{eq:varphi_chiral_transform}
\end{align} 

On the other hand, the expectation value $\langle e^{i\chi}\rangle$ naively
does not transform at all, since at the level of the fundamental fields the
symmetry acts only on $\varphi$. But in fact it turns out that $\langle
e^{i\chi}\rangle \to e^{-2\pi i /p}\, \langle e^{i\chi}\rangle$ when
$\varphi \to \varphi+2\pi/p$. To see this note that the relevant part of the
expectation value is
\begin{align}
\int \mathcal D \hat a\, \mathcal D\chi \,e^{i\chi}\,  e^{-\frac{i}{2\pi}\int\varphi\wedge da-\frac{i}{2\pi}\int\chi\wedge(da-p\, d\hat a)} &= 2\pi \sum_{m\in\ZZ} \delta\left(\frac{1}{2\pi}\int da-1-p\, m \right) e^{-\frac{i}{2\pi}\int\varphi\wedge da}.
\end{align}
We see that in the presence of the insertion $e^{i\chi}$ the constraint on
the topological sectors of $a$ is modified. As a result, when we shift $\varphi
\to \varphi+2\pi/p$, the extra phase $e^{-\frac{i}{p}\int da}$ is a non-trivial
root of unity, and 
\begin{align}
\langle e^{i\chi}\rangle \to e^{-2\pi i /p}\, \langle e^{i\chi}\rangle. 
\label{eq:lambda_chiral_transform_schwinger}
\end{align}
From the point of view of our (naive) definition of the symmetry action on
fields, this is peculiar: we did not assign the field $\chi$ a transformation
under the symmetry, yet correlation functions behave as if $\chi$ were charged.

The action of chiral symmetry on the expectation values of $\varphi$ and $\chi$
found in \eqref{eq:varphi_chiral_transform} and
\eqref{eq:lambda_chiral_transform_schwinger} is consistent with the
identification of $U_k[C] = e^{-i k\int_C (\hat a -
\frac{i}{2p}\star d\varphi) }$ as the charge operators of the $\ZZ_p^{(0)}$ symmetry. One can
easily check that $\langle U_k[C] e^{i\varphi(x)}\rangle = e^{2\pi ik/p}\, \langle
e^{i\varphi(x)}\rangle$ and $\langle U_k[C] e^{i\chi(x)}\rangle = e^{-2\pi i k/p}\,
\langle e^{i\chi(x)}\rangle$ when the closed curve $C$ has linking number $1$
with the point $x$. However, there are no sensible charge operators that act as
$\langle \tilde{U}[C] e^{i\varphi(x)}\rangle = e^{2\pi i/p}\, \langle
e^{i\varphi(x)}\rangle$ and $\langle \tilde{U}[C] e^{i\chi(x)}\rangle = \langle
e^{i\chi(x)}\rangle$. For instance, the natural candidates for
such operators, $\tilde{U}_k[C] = e^{-i k \int_C (\frac{a}{p}-\frac{i}{2p}\star
d\varphi)}$, are not gauge-invariant.

The upshot of this discussion is that to make the action of the chiral symmetry
on fields have the simplest correspondence to its action in correlation
functions, we should define the action of $\ZZ_p^{(0)}$  as 
\begin{equation}
\ZZ_p^{(0)} : \quad \varphi\to \varphi+\frac{2\pi}{p}, \quad \chi\to \chi-\frac{2\pi}{p}.
\end{equation}
With this definition, one can verify that the exponentiated action $e^{-S}$
itself is left invariant thanks to to flux quantization condition for $\hat a$
alone, without integrating out $\chi$ explicitly. This implies that the
expectation values of both $e^{i\varphi}$ and $e^{i\chi}$ transform with the expected
phases.   Passing back to the fermionic formulation of the theory, we see that
the chiral symmetry transformation becomes 
\begin{equation}
    \ZZ_{2p}^{(0)} : \quad \psi \to e^{i\frac{2\pi}{2p}\gamma} \psi, \quad \chi\to \chi-\frac{2\pi}{p}.
\end{equation}

\begin{figure}[h!] 
\centering
\includegraphics[width=0.5\textwidth]{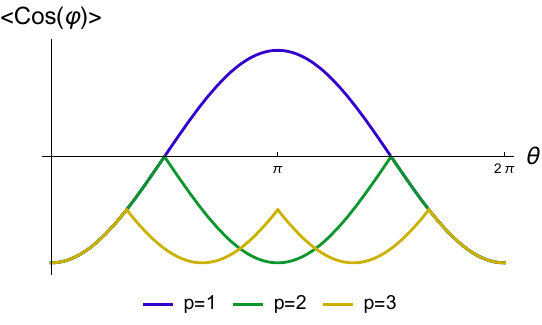}
\caption{The $\theta$-dependence of the chiral condensate $\bar{\psi} \psi \sim
 \cos \varphi$ in the Schwinger model with a very small positive fermion mass $0
 < m \ll e$ and a modified instanton sum characterized by an integer $p$.  The
 unmodified theory has $p=1$. } 
\label{fig:condensate} 
\end{figure}

Finally, let us give an example of how the $\theta$-dependence of local
observables becomes $2\pi/p$-periodic in accordance with the discussion near
Eq.~\eqref{eq:2drestriction}. Consider the expectation value $\langle
\cos\varphi\rangle$ which is proportional to $\langle \bar\psi\psi\rangle$ in
the fermionic variables. Setting $\hat\theta = 0$ without loss of generality,
the expectation value can be computed in the bosonic variables by introducing a
source $J \cos(\varphi)$, minimizing the free energy with respect to $\varphi$,
and taking a derivative with respect to $J$ before sending $J\to0$. As $\theta$
increases from $0$, the value of $\varphi$ minimizing the free energy varies
continuously until $\theta$ crosses $\pi/p$, where it jumps discontinuously as
the minimal-energy universe changes from $k=0$ to $k=1$. This happens again each
time $\theta$ crosses a multiple of $\pi/p$, resulting in the cuspy
$\theta$-dependence of the chiral condensate shown in
Figure~\ref{fig:condensate}. 

\section{Universes and eternal false vacua in 4d}
\label{sec:4d} 

Our goal in this section is to discuss eternal false vacua in 4d gauge theories.
Most of this section is a direct generalization of our discussion in two
dimensions. Our focus will be on $SU(N)$ QCD with $N_f$ fundamental fermions and
a modified instanton sum, which (when the fermions are massive) features eternal
false vacua for generic values of parameters.  This was already noted
in~\cite{Tanizaki:2019rbk} for Yang-Mills theory and one-flavor QCD, and our
discussion in the opening of this section will mostly be a summary of the ideas of~\cite{Seiberg:2010qd,Tanizaki:2019rbk}. 

It is useful to first warm up by recalling the fate of topological charge in QCD
coupled to an axion $\chi$.  The $2\pi$-periodic axion scalar field couples to QCD's
gauge fields through the term 
\begin{align}
    \frac{i}{8\pi^2} \int_{M_4} \chi \wedge \tr F \wedge F
\end{align}
where $F$ is the $SU(N)$ 2-form field strength and $M_4$ is a closed Euclidean spacetime manifold.  Let $\chi_0$ denote the zero
mode of $\chi$. Suppose that the $U(1)_{\text{PQ}}$ Peccei-Quinn symmetry shifting $\chi\to\chi
+ c$ is \emph{only} broken by the coupling to QCD.\footnote{This is a strong
version of the `axion quality' assumption which refers to the constraint that
$U(1)_{\text{PQ}}$-breaking effects in the UV do not spoil $\theta_{\text{eff}}
\approx 0$~\cite{Kamionkowski:1992mf,PhysRevD.46.539}.}   Then QCD instantons generate a potential for the effective theta
parameter $\theta_{\text{eff}} = \chi_0 + \theta$, which is minimized when
$\theta_{\text{eff}} = 0$, rendering the theory CP
invariant~\cite{PhysRevLett.38.1440,PhysRevD.16.1791,PhysRevLett.40.223,PhysRevLett.40.279,Zhitnitsky:1980tq,Kim:1979if,Dine:1981rt,Shifman:1979if}.
In this scenario the zero mode of
$\chi$ only enters linearly in the action, and we
can integrate $\chi_0$ out explicitly, yielding 
\begin{equation}
\int d\chi_0 \, e^{-i\chi_0\int_{M_4}\frac{1}{8\pi^2} \tr\left[ F\wedge F\right]} = 2\pi\, \delta_{Q_T,0}\, ,\label{eq:qcdaxion_constraint} 
\end{equation}
where $Q_T = \int_{M_4}\frac{1}{8\pi^2} \tr\left[ F\wedge F\right]$ is the
topological charge. We see that integration over the zero mode of the QCD axion
implements a constraint that gauge configurations with non-vanishing topological
charge do not contribute to the partition function.  The only way for instantons
to contribute to the partition function is through the effects of 
 instanton-anti-instanton pairs. 

The above global constraint on the QCD partition function arose because we
introduced a new propagating degree of freedom, namely the axion.  Can one
constrain the instanton sum without introducing any new propagating degrees of
freedom?  The answer is yes~\cite{Seiberg:2010qd,Tanizaki:2019rbk}.   
The following action (with a slight abuse of notation) suffices: 
\begin{align}
S = \int_{M_4}\left[\frac{1}{2g^2}\tr\left[F\wedge\star F\right] + \frac{i\theta}{8\pi^2}\tr\left[F\wedge F\right] 
+\sum_{i=1}^{N_f}\left(  \bar\psi_i \gamma^\mu(\partial_\mu-i a_\mu)\psi_i+ m\, \bar\psi_i \psi_i \right)\right] \nonumber \\
+ \int_{M_4}\left[ i\, \chi \wedge \left( \frac{1}{8\pi^2} \tr F \wedge F  - \frac{p}{2\pi} d c^{(3)} \right)
+  \frac{i\hat{\theta}}{2\pi} d c^{(3)}\right]\,, \label{eq:modified_qcd}
\end{align}
where we take $p$ to be a positive integer. The top line is the standard action
for QCD on the spacetime $M_4$: $F = da + i a\wedge a$ where $a$ is the $SU(N)$
gauge field, $\psi_i$ are Dirac fermion fields in the fundamental representation
of the $SU(N)$ gauge group, with the conventional $\theta$ and mass parameter
$m$. The lower line restricts the instanton sum by introducing a 3-form $U(1)$
gauge field $c^{(3)}$ and $2\pi$-periodic scalar Lagrange multiplier
$\chi$.\footnote{We have explicitly included the associated topological angle
$\hat{\theta}$ for the topological charge density $\frac{dc^{(3)}}{2\pi}$ in
addition to the usual QCD $\theta$ parameter. The fact that one can perform
field redefinitions of $\chi$ ensures that  only the linear combination
$p\,\theta + \hat\theta$ can be observable. As usual, when at least one quark
flavor is massless $p\,\theta + \hat\theta$ can be eliminated by a chiral
rotation, but when the quarks are massive the value of $p\,\theta + \hat\theta$
is measurable mod $2\pi$. } The theory enjoys 0-form $SU(N)$ gauge invariance as
well as a 2-form gauge invariance $c^{(3)} \to c^{(3)} + d\lambda^{(2)}$ where
$\lambda^{(2)}$ is a 2-form $U(1)$ gauge function, meaning that $\int d
\lambda^{(2)} \in 2\pi \mathbb{Z}$ on closed 3-manifolds. This implies the flux
quantization condition
\begin{align}
    \frac{1}{2\pi}\int_{M_4} d c^{(3)} \in \mathbb{Z}\,,
\end{align}
where $M_4$ is any closed Euclidean four-manifold, so that $\hat{\theta}$ is a
$2\pi$-periodic parameter. The $\mathbb{Z}^{(1)}_N$ center symmetry
of pure $SU(N)$ YM theory is explicitly broken by the fundamental-representation
fermions. 

The fields $c^{(3)}$ and $\chi$ with the coupling
given in \eqref{eq:modified_qcd} do not
carry any propagating degrees of freedom, and hence do not introduce any new dynamics.\footnote{This assertion relies on
the fact that we did not write kinetic terms for $\chi$ and $c^{(3)}$. Kinetic
terms for 3-form gauge fields are technically irrelevant in four spacetime
dimensions:  the 3-form gauge coupling has mass dimension $2$.  This makes it
consistent to exclude a kinetic term for $c^{(3)}$ in \eqref{eq:modified_qcd}.
But to the extent one can neglect such a kinetic term, the $c^{(3)}$ equation of
motion sets $d\chi = 0$, so it is also consistent to exclude a $\chi$ kinetic
term.  } 
The degrees of freedom described by
\eqref{eq:modified_qcd} are precisely the same as standard QCD.  To see that
introducing these fields restricts the instanton sum, we first note that the
equation of motion for $\chi$ is
\begin{align}
    \frac{1}{8\pi^2} \tr F\wedge F = \frac{p}{2\pi} d c^{(3)} \,.
\end{align}
Equating the integrals of both sides over closed
four-manifolds implies that $SU(N)$ instantons can only contribute to the
partition function if their topological charge is divisible by $p$.  As a
corollary, this immediately implies that the $\theta$ periodicity is $2\pi/p$.
At the same time, the equation of motion of $c^{(3)}$ is
\begin{align}   
d \chi = 0 \,.
\label{eq:constant_chi}
\end{align}
This means that integrating out $c^{(3)}$ sets $\chi$ to be constant, and
recalling the global constraint that $\int d c^{(3)} \in 2\pi \mathbb{Z}$ leads
to the condition $\chi = 2\pi k/p,$ where $k = 0,1, \ldots p-1$. This means that
$e^{i\chi(x)}$ is a topological local operator (i.e. independent of $x$), with
expectation values $\langle e^{i\chi} \rangle = e^{2\pi i k/p}$.  The path
integral over $\chi$ then sums over $k$, so that
\begin{align}
Z_{\rm constrained}(\theta,\hat{\theta}) &= \sum_{k = 0}^{p-1} Z_{\rm original} (\theta+ 2\pi k/p,\hat{\theta}-2\pi k) \nonumber\\
&=\sum_{k = 0}^{p-1} Z_{\rm original} (\theta+ 2\pi k/p,\hat{\theta})
\end{align}
where we used the fact that $\hat{\theta}$ is $2\pi$ periodic. This expression,
which is identical to the 2d expression ~\eqref{eq:decomposition}, gives another
way of seeing that $\theta$ is $2\pi/p$ periodic. It also shows that the
partition function decomposes into universes labelled by $k$, or equivalently,
the expectation values $\langle e^{i\chi}\rangle \in \ZZ_p$.

\subsection{Discrete symmetries and mixed anomaly}
\label{sec:qcd_discrete_symmetries}

Let us now investigate the discrete global symmetries of the theory more
closely. We will discuss the continuous symmetries and their interplay with the
discrete symmetries in the next section.

Let us start with the 0-form internal discrete symmetries. In contrast to
standard massless QCD, when $m=0$ there is an enlarged 0-form discrete chiral
symmetry $\mathbb{Z}_{2N_fp}^{(0)}$ due to the modified instanton sum. As
discussed in Sec.~\ref{sec:modified_sum}, consistency with the expected
transformations of correlation functions requires us to define the
$\mathbb{Z}_{2N_fp}^{(0)}$ symmetry as a combination of chiral rotations of the
fermions and shifts of the Lagrange multiplier field $\chi$. Consider the
$\ZZ_{2N_fp}$ transformations generated by
\begin{align}
  \mathbb{Z}_{2N_fp}^{(0)}:\;\;  \psi_i \to e^{ \frac{2\pi i}{2N_fp} \gamma_5 } \psi_i \,, \qquad \chi \to \chi - \frac{2\pi}{p}
  \label{eq:QCD_chiral_transformation}
\end{align}
The chiral rotation shifts the action by 
\begin{align}
    \delta_{\psi} S =  \frac{2\pi i}{2 N_f p} \times  2N_f \int_{M_4}  \frac{1}{8\pi^2} \tr F \wedge F
    \label{eq:fujikawa}
\end{align}
while the shift of $\chi$ induces a shift of the action by
\begin{align}
    \delta_{\chi} S = - \frac{2\pi i}{p}  
    \int_{M_4} \left[ \frac{1}{8\pi^2} \tr F \wedge F  - \frac{p}{2\pi} d c^{(3)} \right] \,,
    \label{eq:chi_action_transform}
\end{align}
and the invariance of the path integral under any chiral rotation generated by
\eqref{eq:QCD_chiral_transformation} follows from the fact that $\int_{M_4} d
c^{(3)} \in 2\pi \mathbb{Z}$.  The $\mathbb{Z}_2$ subgroup of
$\ZZ_{2N_fp}^{(0)}$ is fermion parity, while the  $\ZZ_{N_f}$ subgroup lies
within $SU(N_f)_A$. Thanks to the modified instanton sum, the discrete part of
the chiral symmetry group is an extension of $\ZZ_{2N_f}$ by $\ZZ_p$ described
by the short exact sequence
\begin{equation}
1 \to \ZZ_{2N_f} \to \ZZ_{2N_fp} \to \ZZ_p \to 1. 
\end{equation}
The standard 't Hooft interaction term $\det \bar{\psi}_i \psi _j$ is only
invariant under $\ZZ_{2N_f}\subset\ZZ_{2N_fp}^{(0)}$, so it cannot be
radiatively generated when $m=0$.  But $e^{i\chi}\det \bar{\psi}_i \psi _j$ is
invariant under $\ZZ_{2N_fp}^{(0)}$, so it can be radiatively generated at
$m=0$.  In any given universe, $\chi$ is a constant, so $e^{i\chi}\det
\bar{\psi}_i \psi _j$ looks just like the standard 't Hooft interaction term.
This makes $\ZZ_{2N_fp}^{(0)}$ a rather peculiar chiral symmetry:  in any given
universe, it does \emph{not} forbid the generation of operators that look
asymmetric.  But $\ZZ_{2N_fp}^{(0)}$ does constrain the form these operators
take in the whole collection of universes, in such a way that the QFT as a whole
is invariant. 

A nonzero common fermion mass $m$ breaks $SU(N_f)_A$ completely
and breaks the discrete chiral symmetry down to fermion number,
$\mathbb{Z}_{2N_fp}^{(0)} \to \ZZ_2$. When $m\not=0$ there remains a vector-like
faithfully acting $U(N_f)/\ZZ_N$ zero-form global symmetry, which will play an important role below.

QCD with a modified instanton sum also has a 3-form global symmetry
$\mathbb{Z}_p^{(3)}$.  One way to see that this must be the case is to observe
that the $c^{(3)}$ gauge field couples with charge $p$ to the (tautologically)
conserved current $\star d \chi$.  If $\chi$ were a propagating axion field,
$\star d \chi$ would be the conserved current for the 2-form $U(1)$ `axion
string' global symmetry.  Gauging an $n$-form $U(1)$ global symmetry with charge
$p$ always leads to a $(n+1)$-form $\mathbb{Z}_p$ global symmetry, as is familiar
from QED coupled to matter with charge $p$.  Indeed, in modified QCD, the topological operator $e^{i \chi}$ can be
interpreted as the symmetry operator for a $3$-form $\ZZ_p^{(3)}$ symmetry
acting on three-dimensional charged objects, which can be viewed as Wilson
`surfaces' $V[M_3] = \exp{[i \int_{M_{3}} c^{(3)}]}$. The existence of the
global symmetry is encoded in the topological correlation function 
\begin{align}
    \langle e^{i \chi(x)} V[M_3] \rangle = e^{\frac{2\pi i}{p} \mathrm{Link}(x,M_3)} \langle V[M_3] \rangle\,,
\end{align}
where $\mathrm{Link}(x,M_3)$ is the linking number of the point $x$ and the
closed 3-manifold $M_3$.  The 3-form global symmetry can also be described in
terms of a transformation acting on the fields: it takes $c^{(3)}
\to c^{(3)}+\lambda^{(3)}$, where $\lambda^{(3)}$ is locally a closed, but not
exact, 3-form, and globally satisfies $\int_{M_3}\lambda^{(3)} \in 2\pi \ZZ/p$.

The fact that the symmetry operator of $\ZZ_p^{(3)}$, namely $e^{i\chi}$, is itself charged under
$\ZZ_{2N_fp}^{(0)}$ implies that there is a mixed anomaly between these two
symmetries. To see this explicitly we turn on background gauge fields for the
$3$-form symmetry.  This amounts to introducing a pair of 3-form and 4-form
background gauge fields $D^{(3)}, D^{(4)}$, and requiring that under a
background 3-form gauge transformation $c^{(3)} \to c^{(3)} + \Lambda^{(3)}$
these background gauge fields transform as
\begin{align}
   D^{(3)} \to D^{(3)} + p\, \Lambda^{(3)} \, \qquad  D^{(4)} \to D^{(4)} + d\Lambda^{(3)}.
\end{align}
We assume that $D^{(3)}$ is a properly normalized $U(1)$ gauge field, so that $\int_{M_4} d D^{(3)} \in 2\pi \ZZ$.  The fields $D_3$ and $D_4$
couple to $c^{(3)}$ via minimal coupling, $p\, c^{(3)} \to p\, c^{(3)} -
D^{(3)}$, and $d c^{(3)} \to d c^{(3)} - D^{(4)}$. To make sure that the $D$
fields describe a $\mathbb{Z}_p$ higher-form gauge theory, we take the $D$
fields to have a BF-type action:
\begin{align}
    \frac{i}{2\pi}    \int_{M_4}  \tilde{\chi} \wedge \left ( p\, D^{(4)} - d D^{(3)}\right)
    + \frac{i\, h}{2\pi}\int_{M_4} D^{(4)}
\label{eq:three_form_constraint}
\end{align}
where $\tilde{\chi}$ is a scalar Lagrange multiplier, and $h = 0, 1,\ldots, p-1$
is a discrete theta parameter.   The equation of motion for
$\tilde{\chi}$ ensures that $\int_{M_4} D^{(4)} \in \frac{2\pi}{p} \ZZ$. At the
same time, the equation of motion for $\chi$ now becomes
\begin{align}
    \frac{1}{8\pi^2} \tr F(a) \wedge F(a) = \frac{1}{2\pi} \left( p\, d c^{(3)} - d D^{(3)} \right)\,.
\label{eq:gauged_EoM_4d}
\end{align}
Applying a chiral transformation shifts the action by 
\begin{align}
    \delta S 
    = i \int_{M_4}  \left( d c^{(3)} -  D^{(4)} \right) \,.
\end{align}
We therefore find the following $\ZZ_{2N_fp}^{(0)}$ transformation of the path
integral in the presence of background fields for $\ZZ_p^{(3)}$: 
\begin{align}
Z(\theta, \hat{\theta}, h; D^{(3)},D^{(4)}) \xrightarrow{\ZZ_{2N_fp}}
    e^{ - i \int D^{(4)}}Z(\theta, \hat{\theta}, h; D^{(3)},D^{(4)}) 
\label{eq:theta_anomaly}
\end{align}
Since $\int D^{(4)} \in \frac{2\pi}{p}\mathbb{Z}$, the phase on the right-hand
side is non-trivial and cannot be eliminated by any choice of local
counter-terms. Thus QCD with a modified instanton sum has a $\mathbb{Z}_p$ 't
Hooft anomaly associated with the quotient $\ZZ_p \simeq \ZZ_{2N_fp}^{(0)}
/\ZZ_{2N_f}$ and the $\ZZ_p^{(3)}$ symmetry.\footnote{It is also possible to
interpret \eqref{eq:theta_anomaly} as an anomaly in the space of couplings (see Ref. 
\cite{Cordova:2019jnf,Cordova:2019uob}) for the $\theta$ parameter.}

Standard assumptions about the behavior of QCD when $N_f < N_f^{*}$, where
$N_f^{*}$ is the lower edge of the conformal window, imply that when $m=0$, the
discrete and continuous chiral symmetries are spontaneously broken on
$\mathbb{R}^4$ when $N_f$ is not too large, with vacua characterized by the
phases of order parameters such as $\sum_i \bar{\psi}_i \psi_i$ and $\det
\bar{\psi}_i \psi_j$.\footnote{We expect that the mixed 't Hooft anomaly is
matched by massless chiral fermions within the conformal window at large $N_f$.}
This means that the Goldstone manifold (which is non-trivial when $N_f >1$) has
$p$ disconnected components corresponding to the $p$ distinct universes.  As
noted above, it is always possible to write the partition function of QCD with a
modified instanton sum as 
\begin{align}
Z(\theta) = \sum_{k = 0}^{p-1} Z\left (\theta +2\pi k/p\right).
\label{eq:partition_sum_QCD}
\end{align}
When $N_f>1$ and the approximate chiral symmetry at small $m$ is spontaneously
broken, one can describe the low-energy dynamics using the chiral field $U =
e^{i \Pi/f_{\pi}}$, where $\Pi$ is a Hermitian matrix and $f_{\pi}$ is the
Nambu-Goldstone boson decay constant.   This must be done separately in each of
the terms in \eqref{eq:partition_sum_QCD}, leading to $p$ disconnected
Goldstone manifolds $U_{k}$,  $k = 0, \ldots, p-1$. 

Finally, we note that the 3-form symmetry is spontaneously broken so long as the
$\ZZ_{2N_fp}^{(0)}$ is not explicitly broken. The argument showing that
$\ZZ_p^{(3)}$ breaks spontaneously is completely analogous to the argument in
Sec.~\ref{sec:chargep_schwinger} showing that the $\ZZ_p^{(1)}$ symmetry breaks
spontaneously in the charge-$p$ Schwinger model. The insertion of the extended
operator $V[M_3]$ in the path integral separates spacetime into two regions.
Inside the four-dimensional subregion bounded by $M_3$, the insertion of
$V[M_3]$ has the same effect as a discrete chiral rotation relative to the
outside region. Therefore, as long as $\ZZ_{2N_fp}^{(0)}$ is not explicitly
broken, the energy cost of inserting $V[M_3]$ does not scale with the spacetime
volume bounded by $M_3$. This is equivalent to saying that the expectation
values of the extended operator $V[M_3]$ follow a `perimeter' law, so the
$\ZZ_p^{(3)}$ symmetry is spontaneously broken.

\subsection{Higher-group symmetry}
\label{sec:qcd_higher_group}

We now turn to the continuous zero-form symmetries of QCD with a modified
instanton sum. It turns out that the 3-form discrete symmetry and the continuous
global symmetry of the theory do not form a direct product.  Instead, they mix
in a non-trivial way, and form a higher-group symmetry structure,
see~\cite{Cordova:2018cvg} for an introduction to higher groups in QFT.   Higher
group symmetries in QFT have been studied in a variety of contexts in recent
work, see
e.g~\cite{BaezLauda,Baez:2004in,Kapustin:2013uxa,Sharpe:2015mja,
Tachikawa:2017gyf,Benini:2018reh,Brennan:2020ehu,Brauner:2020rtz,Wen:2018zux,
Wan:2018bns,Wan:2019soo,Tanizaki:2019rbk,Hidaka:2020iaz,Cordova:2020tij,Hidaka:2020izy,Iqbal:2020lrt}.
In particular, Tanizaki and \"Unsal~\cite{Tanizaki:2019rbk} showed that pure 4d
Yang-Mills theory with a modified instanton sum has a 4-group global symmetry
due to an intertwining between 1-form $\mathbb{Z}_N$ center symmetry and a
$\mathbb{Z}_p$ 3-form global symmetry.   Here we show that $N_f \ge 1$-flavor
QCD with a modified instanton sum also has a 4-group global symmetry, this time
arising due to an intertwining of a continuous zero-form global symmetry with a
$\mathbb{Z}_p$ 3-form global symmetry.    We will show that if one turns on
certain background gauge  fields for the continuous zero-form symmetries, then
one necessarily induces non-trivial fluxes for background gauge fields for the
discrete 3-form global symmetry.

Suppose that all of the flavors of quarks have a common mass $m \neq 0$. Then
the continuous transformations leaving the action invariant are 
\begin{align}
    \frac{SU(N) \times U(N_f)}{\mathbb{Z}_N} 
\end{align}
where the $SU(N)$ factor comes from color gauge transformations, the $U(N_f)$
factor comes from flavor rotations, and the quotient
involves the $\mathbb{Z}_N$ rotations common to both factors.
The physical states are in representations of the faithfully-acting symmetry
group $U(N_f)/\mathbb{Z}_N$. But as observed in Ref.~\cite{Gaiotto:2017tne},
this means that if we turn on an $U(N_f)/\mathbb{Z}_N$ flavor background field
which is not an $U(N_f)$ gauge field, then the dynamical gauge field would be
forced to be in a $PSU(N) = SU(N)/\mathbb{Z}_N$ bundle, rather than in an $SU(N)$
bundle. 

To illustrate how this works, we use the techniques introduced in
\cite{Gaiotto:2014kfa,Kapustin:2014gua}, see also
\cite{Shimizu:2017asf,Cherman:2017tey,Tanizaki:2017bam,Tanizaki:2018wtg,
Anber:2019nze,Kanazawa:2019tnf,Unsal:2020yeh,Anber:2020xfk}.
We will turn a $U(1)_B^{(0)} = U(1)/\mathbb{Z}_N$ baryon number background gauge
field, and will find that in general this leads to the dynamical color gauge
field living in $SU(N)/\mathbb{Z}_N$. One could also probe the theory by turning
on a full non-abelian $U(N_f)/\ZZ_N$ flavor background, but for our purposes it
is sufficient to focus on abelian flavor backgrounds. 

Let us first describe $SU(N)/\mathbb{Z}_N$ gauge
fields.  One way to do this is to couple $SU(N)$ gauge fields to a classical
$\mathbb{Z}_N$ TQFT. The fields of the TQFT are the one-form and two-form
background gauge fields $B^{(2)},B^{(1)}$, with an action
\begin{align}
    S_{\text{TQFT}} = \frac{i}{2\pi} \int_{M_4} \varphi^{(2)} \wedge (N B^{(2)} - d B^{(1)})
\end{align}
where $\varphi^{(2)}$ is a 2-form Lagrange multiplier.  The TQFT
is invariant under $U(1)$ 1-form background gauge transformations $B^{(2)} \to
B^{(2)} + d \lambda^{(1)}$, $B^{(1)} \to B^{(1)} + N \lambda^{(1)}$, with
$\int_{M_2} d \lambda^{(1)} \in 2\pi \mathbb{Z}$ on closed 2-manifolds.  Note that
the flux of $B^{(2)}$ on any closed manifold $M_2$ is fractional, $\int_{M_2}
B^{(2)} \in \frac{2\pi}{N} \mathbb{Z}$. To couple this TQFT to our gauge theory, we 
replace the $SU(N)$ gauge field $a$ by the $U(N)$ gauge field $\tilde{a}$, which
we assign the 1-form gauge transformation $\tilde{a} \to \tilde{a} +
\lambda^{(1)}$, and introduce the extra term in the action
\begin{align}
    S_{U(N)\text{ constraint}} = \frac{i}{2\pi} \int_{M_4} \eta^{(2)} \wedge (\tr \tilde{F} - N B^{(2)})
\end{align}
where $\eta$ is another 2-form Lagrange multiplier and $\tilde{F}$ is
the $U(N)$ field strength.  The $U(N)$ gauge field coupled to $B^{(1)},B^{(2)}$
in the manner given above is a way to write an $SU(N)/\mathbb{Z}_N$ gauge field.
To write the $SU(N)/\mathbb{Z}_N$ gauge field part of the action, we replace
\begin{align}
   \int_{M_4} \left[ \frac{1}{2g^2} \tr F \wedge \star F + \frac{i \theta}{8\pi^2} \tr F \wedge F  \right]\
\end{align}
by
\begin{align}
   \int_{M_4} \left[  \frac{1}{2g^2} \tr (\tilde{F} - B^{(2)} \mathbf{1})\wedge \star (\tilde{F} - B^{(2)} \mathbf{1}) 
    + \frac{i \theta}{8\pi^2} \tr (\tilde{F} - B^{(2)} \mathbf{1}) \wedge (\tilde{F} - B^{(2)} \mathbf{1}) \right]\,
\end{align}
in the Lagrangian.  

We now review the fact that the color gauge field in QCD lies in
$SU(N)/\mathbb{Z}_N$ if we simultaneously turn on a generic background gauge field for
$U(1)_B^{(0)}$.  A $U(1)_B^{(0)}$ gauge field can be described as a quark number gauge
field $A$ coupled to a $\mathbb{Z}_N$ TQFT built out of gauge fields
$C^{(1)},C^{(2)}$ with the constraint $N C^{(2)} = d C^{(1)}$.  The flux of
$C^{(2)}$ is fractional, $\int_{M_2} C^{(2)} \in \frac{2\pi}{N} \mathbb{Z}$.
The fields $A, C^{(1)}, C^{(2)}$ transform under 1-form gauge transformations as
$A \to A + \tilde{\lambda}^{(1)}$, $C^{(1)} \to C^{(1)} + N
\tilde{\lambda}^{(1)}$, $C^{(2)} \to C^{(2)} + d \tilde{\lambda}^{(1)}$.  All
this implies that the quark part of the action of our original $SU(N)$ gauge
theory should be replaced by
\begin{align}
  \int_{M_4} \bar{\psi} \gamma^{\mu}\left[
  \partial_{\mu} 
  - i \left(\tilde{a}  - \frac{B^{(1)}}{N} \right) 
    - i \left(A- \frac{C^{(1)} }{N} \right)  
  \right]\psi\,.
\end{align}
The $1/N$ charges of $\psi$ under the $B^{(1)}$ and $C^{(1)}$ gauge fields mean
that this expression would be inconsistent with charge quantization if only one of
$B^{(1)}$ or $C^{(1)}$ are activated.  But there is no inconsistency as long
as \emph{both} $C^{(1)}$ and $B^{(1)}$ are turned on, with fluxes that obey $\int_{M_2}
B^{(2)} + \int_{M_2} C^{(2)} \in 2\pi \mathbb{Z}$.

Now, in the presence of a nontrivial $U(1)_B^{(0)}$ background, consider the  term in the action
responsible for implementing the constraint on instanton number. Since the dynamical gauge field is in $PSU(N)$, we should replace $F$ with the
$U(N)$ field strength $\tilde{F}$, with $\tr \tilde{F} = N B^{(2)}$ as above. The Lagrange multiplier term becomes
\begin{align}
 \frac{i}{2\pi} \int_{M_4}\left[ \chi \wedge \left( \frac{1}{4\pi} \tr \tilde F \wedge \tilde F  - p\, d c^{(3)} \right)\right]\, .
\end{align}
To make this invariant under background 1-form gauge transformations parametrized by $\lambda^{(1)}$, it is tempting to further replace $\tilde F \to \tilde F - B^{(2)}\mathbf{1}$. However, the result is only gauge invariant up to boundary terms. To make the action completely gauge invariant, we must also turn on a background gauge field for the
$\mathbb{Z}^{(3)}_p$ symmetry. As described in the preceding section, this
involves turning on higher-form gauge fields $D^{(4)}, D^{(3)}$, leading to
\begin{align}
    \frac{i}{2\pi} \int_{M_4}\left[ \chi \wedge \left( \frac{1}{4\pi} \tr \tilde{F} \wedge \tilde{F}  
    - p\, d c^{(3)} + d D^{(3)} \right)\right] \,.
\end{align}
This expression is invariant under 1-form gauge transformations if $D^{(3)}$ transforms as
\begin{align}
    D^{(3)} \to D^{(3)}  - 
    \left( \frac{N}{2\pi} B^{(2)} \wedge \lambda^{(1)} + \frac{N}{4\pi} \lambda^{(1)} \wedge d \lambda^{(1)} \right) \, .
\end{align}
But then we must also replace \eqref{eq:three_form_constraint} with the 1-form gauge-invariant expression
\begin{align}
    \frac{i}{2\pi}    \int_{M_4}  \tilde{\chi} \wedge \left ( p\, D^{(4)} - d D^{(3)} - \frac{N}{4\pi} B^{(2)} \wedge B^{(2)} \right)
    + \frac{i\, h}{2\pi}\int_{M_4} D^{(4)}\,.
\end{align}
As a result we get the relation
\begin{align}
    p\, D^{(4)}  =  d D^{(3)} + \frac{N}{4\pi} B^{(2)} \wedge B^{(2)}
\label{eq:higher_symmetry}
\end{align}
just as in the pure-Yang-Mills case analyzed in \cite{Tanizaki:2019rbk}, so that in general
$
    \int D^{(4)} \in  \frac{2\pi}{N p} \,\mathbb{Z}
$.
Equation \eqref{eq:higher_symmetry} signals that the theory has a higher-group
symmetry: the 0-form and 3-form symmetries do not form a direct product.
Instead, turning on appropriate background gauge fields for the 0-form
$U(1)^{(0)}_B$ global symmetry also requires one to turn on background fields
for the 3-form $\mathbb{Z}^{(3)}_p$ global symmetry.\footnote{When $p=1$, the
$3$-form global symmetry becomes trivial, and the combination $A^{(3)} =
c^{(3)}-D^{(3)}$ defines an ordinary gauge field for the $U(1)^{(2)}$ 2-form
axion string symmetry with field strength $G^{(4)} = dc^{(3)}-D^{(4)}$. Provided
$c^{(3)}$ is interpreted as a non-dynamical background gauge field, the
higher-group structure \eqref{eq:higher_symmetry} then corresponds to the
3-group found in~\cite{Cordova:2020tij}. }

What is the realization of the 4-group symmetry at long distances?   As with any
other symmetry structure, higher-group symmetry can constrain renormalization
group (RG) flows and symmetry-realization patterns. This has been studied in
detail for 2-groups built from continuous 0-form and 1-form symmetries by
Cordova, Dumitrescu and Intriligator\cite{Cordova:2018cvg}. One of the results
of Ref.~\cite{Cordova:2018cvg} is that the 2-group structure implies that it is
inconsistent for the continuous 1-form symmetry to be spontaneously broken if
the continuous 0-form symmetry is not spontaneously broken.
Ref.~\cite{Cordova:2018cvg} also discussed constraints from 2-group symmetry if
the symmetries involved in the 2-group are emergent in the infrared.   In
addition, Ref.~\cite{Cordova:2018cvg} was also able to place bounds on the ratio
of energy scales at which these symmetries emerge, while
Ref.~\cite{Brennan:2020ehu} discussed analogous constraints for some systems
with 3-group symmetries. The heuristic idea behind the observations of
Ref.~\cite{Cordova:2018cvg} is that the zero-form symmetry participating in a
2-group is not a good `subgroup' of the 2-group: turning on generic fluxes for
background gauge fields of the zero-form symmetry induces fluxes for the
background gauge fields of the 1-form symmetry. But fluxes for 1-form symmetry
background fields do not induce fluxes for 0-form symmetry background fields, so
the 1-form symmetry counts as a `good subgroup' of the 2-group symmetry. These
observations motivate the conclusion proved explicitly in
Ref.~\cite{Cordova:2018cvg} that it is only consistent for the 2-group
symmetries discussed in Ref.~\cite{Cordova:2018cvg} to spontaneously break
either to nothing or to the higher-form symmetry group.  The 2-groups discussed
in Ref.~\cite{Cordova:2018cvg}  cannot spontaneously break to the 0-form
symmetry group.

The heuristic picture we summarized above also applies for the 4-group structure
discussed in this paper.  The $U(N_f)/\mathbb{Z}_N$ symmetry group is not a good
`subgroup' of the 4-group symmetry of QCD with a constrained instanton sum in
the sense that fractional background fluxes for $U(N_f)/\mathbb{Z}_N$ induce
non-trivial fluxes for background gauge fields for the $\mathbb{Z}^{(3)}_p$
symmetry.  This observation may make it tempting to guess that the results of
Ref.~\cite{Cordova:2018cvg} apply to the 4-group symmetry discussed here, which
would imply that our 4-group symmetry cannot be spontaneously broken to
$U(N_f)/\mathbb{Z}_N$.

However, a naive generalization along these lines does not work.   From the
discussion at the end of Section~\ref{sec:qcd_discrete_symmetries} we know that
$\mathbb{Z}^{(3)}_p$ is spontaneously broken at $m=0$, while for $m> 0$ it is
not spontaneously broken. When $m\gg \Lambda$, the $U(1)_B^{(0)}$ symmetry will
not be spontaneously broken, and it turns out that the same is true at $m=0$.
To show this, suppose that $\mathcal{O}_f$ is an order parameter for
$U(N_f)/\mathbb{Z}_N$.   If $m =0$ we find
\begin{align}
\langle \mathcal{O}_f \rangle = \frac{\sum_{k = 0}^{p-1}  
\int \mathcal D[\text{fields}]\,e^{-S\left[\mathrm{fields};\, \theta 
= \frac{2\pi k}{ p}\right]} \mathcal{O}_f}{\sum_{k = 0}^{p-1} Z\left(\theta =\frac{2\pi k}{p}\right)} 
=  \frac{\sum_{k = 0}^{p-1} 
Z\left(\theta = \frac{2\pi k}{p}\right) \langle
 \mathcal{O}_f \rangle_{\theta = \frac{2\pi k}{p}}}{\sum_{k = 0}^{p-1} Z\left(\theta = \frac{2\pi k}{p}\right)} = 0
\end{align}
where the expectation value on the left is evaluated in QCD with a constrained
instanton sum, while the partition functions and expectation values in the
middle are evaluated in QCD with an unconstrained instanton sum. The final
equality on the right follows because there is no $\theta$-dependence in the
chiral limit $m \to 0$, so that $\langle \mathcal{O}_f \rangle_{\theta = \frac{2\pi
k}{p}}$ is equal, up to a possible overall phase, to $\langle \mathcal{O}_f
\rangle_{\theta = 0}$. The Vafa-Witten theorem~\cite{Vafa:1983tf} then implies
that the vector-like  $U(N_f)/\mathbb{Z}_N$ cannot be spontaneously broken at
$\theta = 0$, so that $\langle \mathcal{O}_f \rangle_{\theta = 0}$.    So
$U(N_f)/\mathbb{Z}_N$ is not spontaneously broken either at $m=0$ or large $m$,
and it is natural to expect that it is not spontaneously broken for any positive
$m$. Hence, while there is evidence that the 4-group symmetry is not
spontaneously broken at all for $m>0$, when $m=0$ the 4-group symmetry it is
clear that it is spontaneously broken to $U(N_f)/\mathbb{Z}_N$. This is in
contradiction with a naive generalization of the constraints given in
Ref.~\cite{Cordova:2018cvg}.  It would be interesting to do a general study of the
constraints of 4-group symmetries on renormalization group flows, particularly
when the groups involved are discrete.
 
\subsection{Eternal false vacua}
\label{sec:qcd_efv}

When the $\mathbb{Z}^{(3)}_p$ symmetry is exact, the domain walls separating
sectors characterized by different phases $\arg(\det\bar\psi_i\psi_j)$ have
infinite tensions, paralleling what we saw in our 2d example. In infinite volume
the $p$ chiral vacua themselves have an `infinite-fold' degeneracy due to the
spontaneous breaking of $SU(N_f)_A$. The $p$ sectors do not mix even in finite
volume. This implies that even if we increase the temperature of the system
enough to restore $SU(N_f)_A$, $p$ degenerate discrete chiral vacua remain.
Modified QCD thus features `persistent order' in the sense of
\cite{Komargodski:2017dmc}. If we add the term $\det \bar{\psi}_i \psi_j$ to the
Lagrangian and explicitly break $\mathbb{Z}^{(0)}_{2N_fp}\to
\mathbb{Z}_{2N_f}$, the $p$ sectors become non-degenerate, and there is a unique
minimal-energy `ground state.'  But the non-minimal-energy `false vacuum' states
still cannot decay, and each one can still be called a distinct universe.   To
decay, bubbles of the true vacuum would have to nucleate and expand through a
false vacuum.  But in the present context the bubble walls have infinite tension
thanks to the unbroken $\ZZ_p^{(3)}$ symmetry.  So QCD with a modified instanton
sum features eternal false vacua.

The eternal false vacua can be rendered non-eternal by coupling QCD with a
modified instanton sum to dynamical 2-branes with unit charge under
$\mathbb{Z}^{(3)}_p$.  One way to achieve this is to replace the
instanton-number-constraint term in the action
\begin{align}
    \frac{i}{2\pi} \int_{M_4} \chi \wedge \left( \frac{1}{4\pi} \tr F \wedge F - p \, dc^{(3)}\right)
\end{align}
with
\begin{align}
    \frac{i}{2\pi} \int_{M_4} \chi \wedge \left( \frac{1}{4\pi} \tr F \wedge F - \frac{p}{4\pi} 
    \tr F' \wedge F' \right)
    \label{eq:dark_QCD}
\end{align}
where $F'$ is the field strength for some dynamical $SU(N')$ gauge field with
strong scale $\Lambda'$.  Integrating out $\chi$ imposes the constraint that
only $SU(N)$ instantons with topological charge divisible by $p$ contribute to
the path integral.  The $p=1$ version of \eqref{eq:dark_QCD} was recently
considered in Ref.~\cite{Hook:2019qoh} as part of a model to ameliorate the
strong CP axion quality problem.  Equation \eqref{eq:dark_QCD} does not lead to
an exact 3-form symmetry for any $p$.  Consequently, there is no way to forbid
loops of $SU(N')$ gauge bosons from generating a kinetic term for $\chi$.
Moreover, the $\chi \to \chi +2\pi/p$ shift symmetry is still present with
\eqref{eq:dark_QCD}, and in the limit $\Lambda' \gg \Lambda$ we expect an
effective action for the axion-like field $\chi$ of the form
\begin{align}
    S_{\chi} = \frac{1}{2}\int_{M_4} \left[ f^2 |d\chi|^2 + \mu^4 \left(1-\cos(p\chi) + \cdots \right) \right] \, ,
    \label{eq:chi_EFT}
\end{align}
where $f, \mu \sim \Lambda'$.  The field $\chi$ is now dynamical and has $p$
vacua, and one can think of $\frac{i p}{2\pi} \int_{M_4} \chi \wedge d c^{(3)}$
as the long-distance effective field theory description of these vacua as noted
in Ref.~\cite{Tanizaki:2019rbk}.\footnote{See also Ref.~\cite{Anber:2020xfk} for
an interesting discussion of how the 't Hooft anomalies of the $F'$ gauge theory
are matched by domain walls resulting from this kind of effective action.}  So
there is an emergent 3-form symmetry in the infrared when $\Lambda' \gg
\Lambda$.  The 3-form symmetry is not exact because \eqref{eq:chi_EFT} has
dynamical domain-wall excitations with tensions $\sim (\Lambda')^3$ which carry
charge $1$ under $\mathbb{Z}^{(3)}_p$.

We can estimate the decay rate per unit volume for the
discrete vacua using a standard thin-wall bubble nucleation
calculation~\cite{Coleman:1977py}:
\begin{align}
\frac{\Gamma}{V} \sim \mathcal{E} \exp \left[ -\frac{ 27\pi^2}{2} \frac{{T_2}^4}{\mathcal{E} ^3} \right] 
\label{eq:4d_flat_rate}
\end{align}
where $T_2$ is the tension of a unit charge membrane (in the example above $T_2
\sim (\Lambda')^3$) and $\mathcal{E}$ is the difference in vacuum energy
densities between the true and false vacua.  If we break the discrete and continuous chiral symmetries by turning on a flavor symmetric soft mass $m$, $\mathcal{E}$ is set by the physics of QCD and so has a characteristic scale
$m\Lambda_{\rm QCD}^3$, while the membrane tension is a free parameter as far as
QCD is concerned, and could be e.g. $ T_2 \sim M_{\rm Planck}^3$.\footnote{The
estimate in \eqref{eq:4d_flat_rate} is based on a standard flat-space decay rate
calculation, so it is valid provided that the typical bubble size $R \sim
T_2/\mathcal{E}$ is small compared to the inverse Hubble scale $H_0^{-1} \sim
10^{42} \,\textrm{GeV}^{-1}$.  If $\mathcal{E} \sim (1\,\textrm{GeV})^{4}$, then
one should not use \eqref{eq:4d_flat_rate} once $T_2 \gtrsim (10^{-5} M_{\rm
Planck})^3$ given that $M_{\rm Planck} \sim 10^{19}\, \textrm{GeV}$. } If we
write $T_2 = t_b^3 \Lambda_{\rm QCD}^3, \mathcal{E} = \epsilon^4 \Lambda_{\rm
QCD}^4$, then
\begin{align}
    \frac{\Gamma}{V} \sim \exp \left[ - \frac{27\pi^2}{2} 
    \left(\frac{t_b}{\epsilon}\right)^{12}  
    \right]  \label{eq:decayrate} 
\end{align}
so unless one tunes the 2-brane tension parameter $t_b \lesssim \epsilon$, the
decay rate of the discrete false vacua is guaranteed to be spectacularly slow.

\section{Summary}
\label{sec:summary} 

In this paper we have studied quantum field theories exhibiting two somewhat
exotic attributes, both arising from the existence of a $(d-1)$-form global
symmetry. When the symmetry is exact, the path integral decomposes into distinct
universes. When the symmetry is approximate, it is possible to have extremely
long-lived false vacua whose decay rates are set by high-scale physics
controlling the explicit breaking of the higher form symmetry. After a warm up
in quantum mechanics, we examined the phenomena above in generalized versions of
the Schwinger model and four-dimensional QCD. In both cases the global
modifications can be viewed as restricting the sum over instantons to
configurations with topological charge divisible by an integer $p$. The
topological degrees of freedom which lead to the restriction on the topological
charge also give rise to the $(d-1)$-form symmetry responsible for universes and
eternal false vacua. 

Despite the global modifications, which lead to a change in the $\theta$
periodicity, when $\theta = 0$ the theories are locally indistinguishable from
their unmodified counterparts---for example, only extended $(d-1)$-dimensional
objects can probe their different universes.  One key signature of the
modification is the existence of false vacua in generic points in parameter
space. Unlike familiar examples of false vacua, the differences in energy
densities between the false and true vacua can be large, yet the false vacua are
eternal---that is, infinitely long lived---provided the $(d-1)$-form symmetry is
exact. The decay rates of false vacua are then set by the tensions of
$(d-2)$-branes which break the symmetry explicitly. 

In both the Schwinger model and QCD with a modified instanton sum, the lifetimes
of these false vacua can be made arbitrarily large by dialing the hierarchy
between the tensions of the dynamical $(d-2)$-branes and the intrinsic scales of
the theory.  The fact that one can get trapped in a false QCD vacuum for an
extremely long time suggests that the global modification of the instanton sum
is difficult to observe through low-energy measurements near $\theta = 0$, and
should be viewed as an infinite-fold  `ambiguity' in the global structure of
QCD, with a number of parallels to the four-fold global ambiguity of the
Standard Model discussed in Ref.~\cite{Tong:2017oea}. The ambiguity in
Ref.~\cite{Tong:2017oea} arises because of four possible distinct quotients that
can be imposed on the Standard Model gauge group, three of which can lead to
fractionalized topological charges for the $SU(3), SU(2)$ and $U(1)$ gauge
fields.  But it is also possible restrict the topological charges for each
factor of the gauge group individually (or collectively) to be multiples of some
integer(s) with no change to the local dynamics, as explored in this paper for
QCD. We will explore the phenomenological implications of restrictions on the
instanton sum of the Standard Model in future work.

\acknowledgments
We are grateful to Z. Komargodski, M. \"Unsal, and Y. Tanizaki for discussions
that inspired this paper. We also thank M. Shifman and A. Vainshtein for lively
and clarifying discussions about the Schwinger model, and are very grateful to
P. Draper for extensive discussions on the topics in this work and the
suggestion to work out a quantum-mechanical example. A. C. is supported by
start-up funds from the University of Minnesota. T.J. is supported by a UMN CSE
Fellowship.

\bibliographystyle{utphys}
\bibliography{small_circle}

\providecommand{\href}[2]{#2}\begingroup\raggedright\begin{thebibliography}{10}

\bibitem{Kobzarev:1974cp}
I.~Kobzarev, L.~Okun, and M.~Voloshin, ``{Bubbles in Metastable Vacuum},'' {\em
  Sov. J. Nucl. Phys.} {\bfseries 20} (1975) 644--646.

\bibitem{PhysRevD.9.2291}
T.~D. Lee and G.~C. Wick, ``Vacuum stability and vacuum excitation in a spin-0
  field theory,'' \href{http://dx.doi.org/10.1103/PhysRevD.9.2291}{{\em Phys.
  Rev. D} {\bfseries 9} (Apr, 1974) 2291--2316}.
  \url{https://link.aps.org/doi/10.1103/PhysRevD.9.2291}.

\bibitem{PhysRevD.14.3568}
M.~Stone, ``Lifetime and decay of "excited vacuum" states of a field theory
  associated with nonabsolute minima of its effective potential,''
  \href{http://dx.doi.org/10.1103/PhysRevD.14.3568}{{\em Phys. Rev. D}
  {\bfseries 14} (Dec, 1976) 3568--3573}.
  \url{https://link.aps.org/doi/10.1103/PhysRevD.14.3568}.

\bibitem{STONE1977186}
M.~Stone, ``Semiclassical methods for unstable states,''
  \href{http://dx.doi.org/https://doi.org/10.1016/0370-2693(77)90099-5}{{\em
  Physics Letters B} {\bfseries 67} no.~2, (1977) 186 -- 188}.
  \url{http://www.sciencedirect.com/science/article/pii/0370269377900995}.

\bibitem{Coleman:1977py}
S.~R. Coleman, ``{The Fate of the False Vacuum. 1. Semiclassical Theory},''
  \href{http://dx.doi.org/10.1103/PhysRevD.16.1248}{{\em Phys. Rev. D}
  {\bfseries 15} (1977) 2929--2936}. [Erratum: Phys.Rev.D 16, 1248 (1977)].

\bibitem{Callan:1977pt}
J.~Callan, Curtis~G. and S.~R. Coleman, ``{The Fate of the False Vacuum. 2.
  First Quantum Corrections},''
  \href{http://dx.doi.org/10.1103/PhysRevD.16.1762}{{\em Phys. Rev. D}
  {\bfseries 16} (1977) 1762--1768}.

\bibitem{ZoharPrivateCommunications}
Z.~Komargodski {\em private communication} (October, 2019) .

\bibitem{Tanizaki:2019rbk}
Y.~Tanizaki and M.~\"Unsal, ``{Modified instanton sum in QCD and
  higher-groups},'' \href{http://dx.doi.org/10.1007/JHEP03(2020)123}{{\em JHEP}
  {\bfseries 03} (2020) 123}, \href{http://arxiv.org/abs/1912.01033}{{\ttfamily
  arXiv:1912.01033 [hep-th]}}.

\bibitem{Komargodski:2020mxz}
Z.~Komargodski, K.~Ohmori, K.~Roumpedakis, and S.~Seifnashri, ``{Symmetries and
  Strings of Adjoint QCD${}_2$},''
  \href{http://arxiv.org/abs/2008.07567}{{\ttfamily arXiv:2008.07567
  [hep-th]}}.

\bibitem{Palti:2019pca}
E.~Palti, ``{The Swampland: Introduction and Review},''
  \href{http://dx.doi.org/10.1002/prop.201900037}{{\em Fortsch. Phys.}
  {\bfseries 67} no.~6, (2019) 1900037},
  \href{http://arxiv.org/abs/1903.06239}{{\ttfamily arXiv:1903.06239
  [hep-th]}}.

\bibitem{Gaiotto:2017yup}
D.~Gaiotto, A.~Kapustin, Z.~Komargodski, and N.~Seiberg, ``{Theta, time
  reversal, and temperature},''
  \href{http://dx.doi.org/10.1007/JHEP05(2017)091}{{\em JHEP} {\bfseries 05}
  (2017) 091},
\href{http://arxiv.org/abs/1703.00501}{{\ttfamily arXiv:1703.00501 [hep-th]}}.

\bibitem{Kikuchi:2017pcp}
Y.~Kikuchi and Y.~Tanizaki, ``{Global inconsistency, 't~Hooft anomaly, and
  level crossing in quantum mechanics},''
\href{http://arxiv.org/abs/1708.01962}{{\ttfamily arXiv:1708.01962 [hep-th]}}.

\bibitem{Aitken:2018kky}
K.~Aitken, A.~Cherman, and M.~{\"U}nsal, ``{Dihedral symmetry in $SU(N)$
  Yang-Mills theory},''
\href{http://arxiv.org/abs/1804.05845}{{\ttfamily arXiv:1804.05845 [hep-th]}}.

\bibitem{Komargodski:2017dmc}
Z.~Komargodski, A.~Sharon, R.~Thorngren, and X.~Zhou, ``{Comments on Abelian
  Higgs models and persistent order},''
\href{http://arxiv.org/abs/1705.04786}{{\ttfamily arXiv:1705.04786 [hep-th]}}.

\bibitem{Gaiotto:2014kfa}
D.~Gaiotto, A.~Kapustin, N.~Seiberg, and B.~Willett, ``{Generalized global
  symmetries},'' \href{http://dx.doi.org/10.1007/JHEP02(2015)172}{{\em JHEP}
  {\bfseries 02} (2015) 172},
\href{http://arxiv.org/abs/1412.5148}{{\ttfamily arXiv:1412.5148 [hep-th]}}.

\bibitem{Anber:2018jdf}
M.~M. Anber and E.~Poppitz, ``{Anomaly matching, (axial) Schwinger models, and
  high-T super Yang-Mills domain walls},''
  \href{http://dx.doi.org/10.1007/JHEP09(2018)076}{{\em JHEP} {\bfseries 09}
  (2018) 076}, \href{http://arxiv.org/abs/1807.00093}{{\ttfamily
  arXiv:1807.00093 [hep-th]}}.

\bibitem{Anber:2018xek}
M.~M. Anber and E.~Poppitz, ``{Domain walls in high-T SU(N) super Yang-Mills
  theory and QCD(adj)},'' \href{http://dx.doi.org/10.1007/JHEP05(2019)151}{{\em
  JHEP} {\bfseries 05} (2019) 151},
  \href{http://arxiv.org/abs/1811.10642}{{\ttfamily arXiv:1811.10642
  [hep-th]}}.

\bibitem{Misumi:2019dwq}
T.~Misumi, Y.~Tanizaki, and M.~{\"U}nsal, ``{Fractional $\theta$ angle, 't
  Hooft anomaly, and quantum instantons in charge-$q$ multi-flavor Schwinger
  model},'' \href{http://dx.doi.org/10.1007/JHEP07(2019)018}{{\em JHEP}
  {\bfseries 07} (2019) 018}, \href{http://arxiv.org/abs/1905.05781}{{\ttfamily
  arXiv:1905.05781 [hep-th]}}.

\bibitem{Armoni:2018bga}
A.~Armoni and S.~Sugimoto, ``{Vacuum structure of charge k two-dimensional QED
  and dynamics of an anti D-string near an O1--plane},''
  \href{http://dx.doi.org/10.1007/JHEP03(2019)175}{{\em JHEP} {\bfseries 03}
  (2019) 175}, \href{http://arxiv.org/abs/1812.10064}{{\ttfamily
  arXiv:1812.10064 [hep-th]}}.

\bibitem{Lawrence:2012ua}
A.~Lawrence, ``{\textbackslash{}theta-angle monodromy in two dimensions},''
  \href{http://dx.doi.org/10.1103/PhysRevD.85.105029}{{\em Phys. Rev. D}
  {\bfseries 85} (2012) 105029},
  \href{http://arxiv.org/abs/1203.6656}{{\ttfamily arXiv:1203.6656 [hep-th]}}.

\bibitem{PhysRevD.11.2088}
S.~Coleman, ``Quantum sine-gordon equation as the massive thirring model,''
  \href{http://dx.doi.org/10.1103/PhysRevD.11.2088}{{\em Phys. Rev. D}
  {\bfseries 11} (Apr, 1975) 2088--2097}.
  \url{https://link.aps.org/doi/10.1103/PhysRevD.11.2088}.

\bibitem{Coleman:1976uz}
S.~R. Coleman, ``{More About the Massive Schwinger Model},''
  \href{http://dx.doi.org/10.1016/0003-4916(76)90280-3}{{\em Annals Phys.}
  {\bfseries 101} (1976) 239}.

\bibitem{Pantev:2005rh}
T.~Pantev and E.~Sharpe, ``{Notes on gauging noneffective group actions},''
  \href{http://arxiv.org/abs/hep-th/0502027}{{\ttfamily arXiv:hep-th/0502027}}.

\bibitem{Seiberg:2010qd}
N.~Seiberg, ``{Modifying the Sum Over Topological Sectors and Constraints on
  Supergravity},'' \href{http://dx.doi.org/10.1007/JHEP07(2010)070}{{\em JHEP}
  {\bfseries 07} (2010) 070}, \href{http://arxiv.org/abs/1005.0002}{{\ttfamily
  arXiv:1005.0002 [hep-th]}}.

\bibitem{Hellerman:2006zs}
S.~Hellerman, A.~Henriques, T.~Pantev, E.~Sharpe, and M.~Ando, ``{Cluster
  decomposition, T-duality, and gerby CFT's},''
  \href{http://dx.doi.org/10.4310/ATMP.2007.v11.n5.a2}{{\em Adv. Theor. Math.
  Phys.} {\bfseries 11} no.~5, (2007) 751--818},
  \href{http://arxiv.org/abs/hep-th/0606034}{{\ttfamily arXiv:hep-th/0606034}}.

\bibitem{Kamionkowski:1992mf}
M.~Kamionkowski and J.~March-Russell, ``{Planck scale physics and the
  Peccei-Quinn mechanism},''
  \href{http://dx.doi.org/10.1016/0370-2693(92)90492-M}{{\em Phys. Lett. B}
  {\bfseries 282} (1992) 137--141},
  \href{http://arxiv.org/abs/hep-th/9202003}{{\ttfamily arXiv:hep-th/9202003}}.

\bibitem{PhysRevD.46.539}
S.~M. Barr and D.~Seckel, ``Planck-scale corrections to axion models,''
  \href{http://dx.doi.org/10.1103/PhysRevD.46.539}{{\em Phys. Rev. D}
  {\bfseries 46} (Jul, 1992) 539--549}.
  \url{https://link.aps.org/doi/10.1103/PhysRevD.46.539}.

\bibitem{PhysRevLett.38.1440}
R.~D. Peccei and H.~R. Quinn, ``$\mathrm{CP}$ conservation in the presence of
  pseudoparticles,'' \href{http://dx.doi.org/10.1103/PhysRevLett.38.1440}{{\em
  Phys. Rev. Lett.} {\bfseries 38} (Jun, 1977) 1440--1443}.
  \url{https://link.aps.org/doi/10.1103/PhysRevLett.38.1440}.

\bibitem{PhysRevD.16.1791}
R.~D. Peccei and H.~R. Quinn, ``Constraints imposed by $\mathrm{CP}$
  conservation in the presence of pseudoparticles,''
  \href{http://dx.doi.org/10.1103/PhysRevD.16.1791}{{\em Phys. Rev. D}
  {\bfseries 16} (Sep, 1977) 1791--1797}.
  \url{https://link.aps.org/doi/10.1103/PhysRevD.16.1791}.

\bibitem{PhysRevLett.40.223}
S.~Weinberg, ``A new light boson?''
  \href{http://dx.doi.org/10.1103/PhysRevLett.40.223}{{\em Phys. Rev. Lett.}
  {\bfseries 40} (Jan, 1978) 223--226}.
  \url{https://link.aps.org/doi/10.1103/PhysRevLett.40.223}.

\bibitem{PhysRevLett.40.279}
F.~Wilczek, ``Problem of strong $p$ and $t$ invariance in the presence of
  instantons,'' \href{http://dx.doi.org/10.1103/PhysRevLett.40.279}{{\em Phys.
  Rev. Lett.} {\bfseries 40} (Jan, 1978) 279--282}.
  \url{https://link.aps.org/doi/10.1103/PhysRevLett.40.279}.

\bibitem{Zhitnitsky:1980tq}
A.~Zhitnitsky, ``{On Possible Suppression of the Axion Hadron Interactions. (In
  Russian)},'' {\em Sov. J. Nucl. Phys.} {\bfseries 31} (1980) 260.

\bibitem{Kim:1979if}
J.~E. Kim, ``{Weak Interaction Singlet and Strong CP Invariance},''
  \href{http://dx.doi.org/10.1103/PhysRevLett.43.103}{{\em Phys. Rev. Lett.}
  {\bfseries 43} (1979) 103}.

\bibitem{Dine:1981rt}
M.~Dine, W.~Fischler, and M.~Srednicki, ``{A Simple Solution to the Strong CP
  Problem with a Harmless Axion},''
  \href{http://dx.doi.org/10.1016/0370-2693(81)90590-6}{{\em Phys. Lett. B}
  {\bfseries 104} (1981) 199--202}.

\bibitem{Shifman:1979if}
M.~A. Shifman, A.~Vainshtein, and V.~I. Zakharov, ``{Can Confinement Ensure
  Natural CP Invariance of Strong Interactions?},''
  \href{http://dx.doi.org/10.1016/0550-3213(80)90209-6}{{\em Nucl. Phys. B}
  {\bfseries 166} (1980) 493--506}.

\bibitem{Cordova:2019jnf}
C.~C\'ordova, D.~S. Freed, H.~T. Lam, and N.~Seiberg, ``{Anomalies in the Space
  of Coupling Constants and Their Dynamical Applications I},''
  \href{http://dx.doi.org/10.21468/SciPostPhys.8.1.001}{{\em SciPost Phys.}
  {\bfseries 8} no.~1, (2020) 001},
  \href{http://arxiv.org/abs/1905.09315}{{\ttfamily arXiv:1905.09315
  [hep-th]}}.

\bibitem{Cordova:2019uob}
C.~C\'ordova, D.~S. Freed, H.~T. Lam, and N.~Seiberg, ``{Anomalies in the Space
  of Coupling Constants and Their Dynamical Applications II},''
  \href{http://dx.doi.org/10.21468/SciPostPhys.8.1.002}{{\em SciPost Phys.}
  {\bfseries 8} no.~1, (2020) 002},
  \href{http://arxiv.org/abs/1905.13361}{{\ttfamily arXiv:1905.13361
  [hep-th]}}.

\bibitem{Cordova:2018cvg}
C.~C{\'o}rdova, T.~T. Dumitrescu, and K.~Intriligator, ``{Exploring 2-Group
  global symmetries},''
\href{http://arxiv.org/abs/1802.04790}{{\ttfamily arXiv:1802.04790 [hep-th]}}.

\bibitem{BaezLauda}
J.~C. {Baez} and A.~D. {Lauda}, ``{Higher-dimensional algebra V: 2-Groups},''
  {\em ArXiv Mathematics e-prints} (July, 2003) ,
  \href{http://arxiv.org/abs/math/0307200}{{\ttfamily math/0307200}}.

\bibitem{Baez:2004in}
J.~Baez and U.~Schreiber, ``{Higher gauge theory: 2-connections on
  2-bundles},'' \href{http://arxiv.org/abs/hep-th/0412325}{{\ttfamily
  arXiv:hep-th/0412325}}.

\bibitem{Kapustin:2013uxa}
A.~Kapustin and R.~Thorngren, ``{Higher symmetry and gapped phases of gauge
  theories},''
\href{http://arxiv.org/abs/1309.4721}{{\ttfamily arXiv:1309.4721 [hep-th]}}.

\bibitem{Sharpe:2015mja}
E.~Sharpe, ``{Notes on generalized global symmetries in QFT},''
  \href{http://dx.doi.org/10.1002/prop.201500048}{{\em Fortsch. Phys.}
  {\bfseries 63} (2015) 659--682},
  \href{http://arxiv.org/abs/1508.04770}{{\ttfamily arXiv:1508.04770
  [hep-th]}}.

\bibitem{Tachikawa:2017gyf}
Y.~Tachikawa, ``{On gauging finite subgroups},''
  \href{http://dx.doi.org/10.21468/SciPostPhys.8.1.015}{{\em SciPost Phys.}
  {\bfseries 8} no.~1, (2020) 015},
  \href{http://arxiv.org/abs/1712.09542}{{\ttfamily arXiv:1712.09542
  [hep-th]}}.

\bibitem{Benini:2018reh}
F.~Benini, C.~Cordova, and P.-S. Hsin, ``{On 2-Group global symmetries and
  their anomalies},''
\href{http://arxiv.org/abs/1803.09336}{{\ttfamily arXiv:1803.09336 [hep-th]}}.

\bibitem{Brennan:2020ehu}
T.~D. Brennan and C.~Cordova, ``{Axions, Higher-Groups, and Emergent
  Symmetry},'' \href{http://arxiv.org/abs/2011.09600}{{\ttfamily
  arXiv:2011.09600 [hep-th]}}.

\bibitem{Brauner:2020rtz}
T.~Brauner, ``{Field theories with higher-group symmetry from composite
  currents},'' \href{http://arxiv.org/abs/2012.00051}{{\ttfamily
  arXiv:2012.00051 [hep-th]}}.

\bibitem{Wen:2018zux}
X.-G. Wen, ``{Emergent anomalous higher symmetries from topological order and
  from dynamical electromagnetic field in condensed matter systems},''
  \href{http://dx.doi.org/10.1103/PhysRevB.99.205139}{{\em Phys. Rev. B}
  {\bfseries 99} no.~20, (2019) 205139},
  \href{http://arxiv.org/abs/1812.02517}{{\ttfamily arXiv:1812.02517
  [cond-mat.str-el]}}.

\bibitem{Wan:2018bns}
Z.~Wan and J.~Wang, ``{Higher anomalies, higher symmetries, and cobordisms I:
  classification of higher-symmetry-protected topological states and their
  boundary fermionic/bosonic anomalies via a generalized cobordism theory},''
  \href{http://dx.doi.org/10.4310/AMSA.2019.v4.n2.a2}{{\em Ann. Math. Sci.
  Appl.} {\bfseries 4} no.~2, (2019) 107--311},
  \href{http://arxiv.org/abs/1812.11967}{{\ttfamily arXiv:1812.11967
  [hep-th]}}.

\bibitem{Wan:2019soo}
Z.~Wan, J.~Wang, and Y.~Zheng, ``{Higher anomalies, higher symmetries, and
  cobordisms II: Lorentz symmetry extension and enriched bosonic / fermionic
  quantum gauge theory},''
  \href{http://dx.doi.org/10.4310/AMSA.2020.v5.n2.a2}{{\em Ann. Math. Sci.
  Appl.} {\bfseries 05} no.~2, (2020) 171--257},
  \href{http://arxiv.org/abs/1912.13504}{{\ttfamily arXiv:1912.13504
  [hep-th]}}.

\bibitem{Hidaka:2020iaz}
Y.~Hidaka, M.~Nitta, and R.~Yokokura, ``{Higher-form symmetries and 3-group in
  axion electrodynamics},''
  \href{http://dx.doi.org/10.1016/j.physletb.2020.135672}{{\em Phys. Lett. B}
  {\bfseries 808} (2020) 135672},
  \href{http://arxiv.org/abs/2006.12532}{{\ttfamily arXiv:2006.12532
  [hep-th]}}.

\bibitem{Cordova:2020tij}
C.~Cordova, T.~T. Dumitrescu, and K.~Intriligator, ``{2-Group Global Symmetries
  and Anomalies in Six-Dimensional Quantum Field Theories},''
  \href{http://arxiv.org/abs/2009.00138}{{\ttfamily arXiv:2009.00138
  [hep-th]}}.

\bibitem{Hidaka:2020izy}
Y.~Hidaka, M.~Nitta, and R.~Yokokura, ``{Global 3-group symmetry and 't Hooft
  anomalies in axion electrodynamics},''
  \href{http://arxiv.org/abs/2009.14368}{{\ttfamily arXiv:2009.14368
  [hep-th]}}.

\bibitem{Iqbal:2020lrt}
N.~Iqbal and N.~Poovuttikul, ``{2-group global symmetries, hydrodynamics and
  holography},'' \href{http://arxiv.org/abs/2010.00320}{{\ttfamily
  arXiv:2010.00320 [hep-th]}}.

\bibitem{Gaiotto:2017tne}
D.~Gaiotto, Z.~Komargodski, and N.~Seiberg, ``{Time-reversal breaking in
  QCD$_4$, walls, and dualities in 2+1 dimensions},''
\href{http://arxiv.org/abs/1708.06806}{{\ttfamily arXiv:1708.06806 [hep-th]}}.

\bibitem{Kapustin:2014gua}
A.~Kapustin and N.~Seiberg, ``{Coupling a QFT to a TQFT and duality},''
  \href{http://dx.doi.org/10.1007/JHEP04(2014)001}{{\em JHEP} {\bfseries 04}
  (2014) 001},
\href{http://arxiv.org/abs/1401.0740}{{\ttfamily arXiv:1401.0740 [hep-th]}}.

\bibitem{Shimizu:2017asf}
H.~Shimizu and K.~Yonekura, ``{Anomaly constraints on deconfinement and chiral
  phase transition},''
\href{http://arxiv.org/abs/1706.06104}{{\ttfamily arXiv:1706.06104 [hep-th]}}.

\bibitem{Cherman:2017tey}
A.~Cherman, S.~Sen, M.~Unsal, M.~L. Wagman, and L.~G. Yaffe, ``{Order
  parameters and color-flavor center symmetry in QCD},''
  \href{http://dx.doi.org/10.1103/PhysRevLett.119.222001}{{\em Phys. Rev.
  Lett.} {\bfseries 119} no.~22, (2017) 222001},
\href{http://arxiv.org/abs/1706.05385}{{\ttfamily arXiv:1706.05385 [hep-th]}}.

\bibitem{Tanizaki:2017bam}
Y.~Tanizaki and Y.~Kikuchi, ``{Vacuum structure of bifundamental gauge theories
  at finite topological angles},''
  \href{http://dx.doi.org/10.1007/JHEP06(2017)102}{{\em JHEP} {\bfseries 06}
  (2017) 102},
\href{http://arxiv.org/abs/1705.01949}{{\ttfamily arXiv:1705.01949 [hep-th]}}.

\bibitem{Tanizaki:2018wtg}
Y.~Tanizaki, ``{Anomaly constraint on massless QCD and the role of Skyrmions in
  chiral symmetry breaking},''
  \href{http://dx.doi.org/10.1007/JHEP08(2018)171}{{\em JHEP} {\bfseries 08}
  (2018) 171}, \href{http://arxiv.org/abs/1807.07666}{{\ttfamily
  arXiv:1807.07666 [hep-th]}}.

\bibitem{Anber:2019nze}
M.~M. Anber and E.~Poppitz, ``{On the baryon-color-flavor (BCF) anomaly in
  vector-like theories},''
  \href{http://dx.doi.org/10.1007/JHEP11(2019)063}{{\em JHEP} {\bfseries 11}
  (2019) 063}, \href{http://arxiv.org/abs/1909.09027}{{\ttfamily
  arXiv:1909.09027 [hep-th]}}.

\bibitem{Kanazawa:2019tnf}
T.~Kanazawa and M.~\"Unsal, ``{Quantum distillation in QCD},''
  \href{http://dx.doi.org/10.1103/PhysRevD.102.034013}{{\em Phys. Rev. D}
  {\bfseries 102} no.~3, (2020) 034013},
  \href{http://arxiv.org/abs/1909.05222}{{\ttfamily arXiv:1909.05222
  [hep-th]}}.

\bibitem{Unsal:2020yeh}
M.~\"Unsal, ``{Strongly coupled QFT dynamics via TQFT coupling},''
  \href{http://arxiv.org/abs/2007.03880}{{\ttfamily arXiv:2007.03880
  [hep-th]}}.

\bibitem{Anber:2020xfk}
M.~M. Anber and E.~Poppitz, ``{Deconfinement on axion domain walls},''
  \href{http://dx.doi.org/10.1007/JHEP03(2020)124}{{\em JHEP} {\bfseries 03}
  (2020) 124}, \href{http://arxiv.org/abs/2001.03631}{{\ttfamily
  arXiv:2001.03631 [hep-th]}}.

\bibitem{Vafa:1983tf}
C.~Vafa and E.~Witten, ``{Restrictions on Symmetry Breaking in Vector-Like
  Gauge Theories},'' \href{http://dx.doi.org/10.1016/0550-3213(84)90230-X}{{\em
  Nucl. Phys. B} {\bfseries 234} (1984) 173--188}.

\bibitem{Hook:2019qoh}
A.~Hook, S.~Kumar, Z.~Liu, and R.~Sundrum, ``{High Quality QCD Axion and the
  LHC},'' \href{http://dx.doi.org/10.1103/PhysRevLett.124.221801}{{\em Phys.
  Rev. Lett.} {\bfseries 124} no.~22, (2020) 221801},
  \href{http://arxiv.org/abs/1911.12364}{{\ttfamily arXiv:1911.12364
  [hep-ph]}}.

\bibitem{Tong:2017oea}
D.~Tong, ``{Line Operators in the Standard Model},''
  \href{http://dx.doi.org/10.1007/JHEP07(2017)104}{{\em JHEP} {\bfseries 07}
  (2017) 104}, \href{http://arxiv.org/abs/1705.01853}{{\ttfamily
  arXiv:1705.01853 [hep-th]}}.

\end{thebibliography}\endgroup

 \end{document}